# Comparative Evaluation of Encapsulation Methods for Endohedral Doping of Single-Wall Carbon Nanotubes

*Cristian M. Borja-Peña[1,2,3], Martin Magg[4], Hussein Slim[5], Yamaldi M. Bakary[5], Timm Swoboda[6], Cedric Desroches[5], Denys Kurylo[3], Kazuhiro Yanagi[7], Marianna Sledzinska[6], Benjamin S. Flavel[4], Salomé Forel[5,*], Wim Wenseleers[3], Sofie Cambré[1,2,*]*

[1] University of Antwerp, Physics Department, Theory and Spectroscopy of Molecules and Materials (TSM[2]).

[2] CASCH Center of Excellence, University of Antwerp, Antwerp, Belgium.

[3] University of Antwerp, Physics Department, Nanostructured and Organic Optical and Electronic Materials (NANOrOPT).

[4] Institute of Nanotechnology, Karlsruhe Institute of Technology, Kaiserstraße 12, 76131 Karlsruhe, Germany.

[5] Claude Bernard University Lyon 1, Laboratoire des Multimatériaux et Interfaces (LMI).

[6] Catalan Institute of Nanoscience and Nanotechnology (ICN2), CSIC and BIST, Barcelona, 08193, Spain.

[7] Tokyo Metropolitan University, Department of Physics, 1-1 Minami-Osawa, Hachioji, Tokyo, Japan 1920397.

ABSTRACT.

Single-wall carbon nanotubes (SWCNTs) are promising building blocks for nanoelectronic and optoelectronic devices, yet reliable and stable doping, particularly n-type, remains challenging due to strong environmental sensitivity and competing extrinsic effects. Encapsulation of charge-transfer molecules within the SWCNT cavity offers a promising route to stable doping while preserving the nanotube's outer surface for subsequent processing. Here, we systematically investigate the filling of arc-discharge SWCNTs with the electron donor tetrathiafulvalene and electron acceptor tetracyanoquinodimethane, comparing different methods for filling, including melt filling, solution reflux, and vacuum-phase sublimation. We follow the entire processing workflow from raw, unfilled powders to aqueous dispersions and employ density gradient ultracentrifugation to separate filled from empty nanotubes as well as metallic from semiconducting ones. Encapsulation efficiency and electronic modification are assessed using absorption spectroscopy, resonant Raman scattering, thermogravimetric analysis, X-ray photoelectron spectroscopy and electron paramagnetic resonance. Finally, we introduce a complementary vacuum-phase method that removes externally adsorbed molecules without extensive solvent washing, enabling cleaner encapsulated systems.

## 1. INTRODUCTION

Single-wall carbon nanotubes (SWCNTs) have emerged as one of the most promising one-dimensional materials for next-generation nanoelectronic and optoelectronic technologies.[1,2] Their unique atomic structure, conceptually a graphene sheet rolled into a seamless cylinder, gives rise to an exceptional combination of electronic, optical, and mechanical properties. Semiconducting SWCNTs, in particular, display high carrier mobilities[3] and low inter-nanotube contact resistance,[4] making them promising candidates for field-effect transistors (FETs) with a high on/off ratio.[5–7]

To create logic devices, SWCNT-FETs need to be made from both p- and n-type SWCNTs rather than unipolar devices.[8–10] However, SWCNTs typically exhibit a strong tendency for intrinsic p-type behavior, primarily due to unintentional doping by atmospheric oxidants (such as oxygen), which withdraw electrons from the SWCNT surface, thereby shifting the Fermi level toward the valence band.[11] This makes p-type conduction the default state of SWCNT-FETs. In contrast, achieving n-type doping in CNTs remains a challenge. Although electron donors can convert CNTs to n-type, these states are typically unstable under ambient conditions because oxygen readily re-oxidizes the nanotubes, unless the outer surface is sufficiently protected.[12]

Doping can be achieved in various manners, either by incorporating hetero-atoms into the SWCNT lattice,[13] electrochemically,[14,15] by encapsulating molecules in their interior hollow cavity,[16–19] or by adsorbing molecules on their surface.[20–22] In this work, we focus on the encapsulation of two well-known electron donor and electron acceptor molecules, tetrathiafulvalene (TTF) and tetracyanoquinodimethane (TCNQ) respectively, which are frequently employed for both external adsorption[23] and encapsulation[16] doping of SWCNTs. In particular, the latter is of interest to increase the stability of the doping, as encapsulated molecules are protected from photochemical attack and environmental interactions.[24] Moreover, encapsulation keeps the external SWCNT structure intact and accessible for further processing, such as required when assembling SWCNT transistors with DNA origami[25] or for further structure sorting by density gradient ultracentrifugation[26,27], polymer wrapping[28] or aqueous two-phase extraction (ATPE).[29,30]

A wide range of strategies exists for introducing foreign molecules into the hollow core of SWCNTs, with the choice of method largely dictated by the physical and chemical characteristics of the filling species, as recently reviewed.[31] Filling procedures can broadly be categorized into solution-based, melt-phase, and vapor-phase (sublimation) methods. In solution-based filling, the pre-opened nanotubes[32] are immersed in a solvent containing the dissolved compound.[33,34] Alternatively, when the filling compound is liquid at relatively low temperature, it may be introduced from the neat liquid (molten) phase,[35] enabling high filling yields and the formation of long, continuous nanocrystals in case the target substance is solid at room temperature.[36] For materials with high melting points or limited solubility but sufficient stability, vapor-phase methods (sublimation filling), provide an effective route, relying on the diffusion of vaporized species into the nanotube cavity.[16]

Although numerous studies have reported encapsulation of molecules within CNTs,[31] a detailed and systematic characterization of the specific filling procedures remains elusive. Typically, a method is chosen on a case-by-case basis depending on the properties of the encapsulating molecules. Therefore, most evaluations on filling efficiency have remained largely qualitative, underscoring the need for more systematic and quantitative studies in this area.

In the context of doping characterization, one of the central challenges lies in disentangling the numerous competing effects that simultaneously influence the optical and electronic response of CNT samples. For example, doping induced by externally adsorbed oxygen complicates the interpretation of doping from the encapsulating molecules. Additionally, strain[37], bundling[38], doping[14,39,40], and external adsorption of molecules[41] can produce similar signatures in Raman spectroscopy, a common technique to investigate the doping, making it difficult to attribute observed spectral changes to a single physical origin. Moreover, optical signatures of different SWCNT diameters react differently to intentional doping.[14] The situation is further complicated by strong dependencies on sample morphology, for example, isolated versus bundled SWCNTs, powders or individually-solubilized SWCNTs in solution, or film-deposited networks, each interact differently, leading to distinct spectroscopic responses and complicating cross-sample comparisons.

In this work, we study the encapsulation of TTF and TCNQ within the hollow core of arc-discharge SWCNTs using three different filling approaches: melt filling, reflux-based solution filling, and vacuum-phase sublimation. We compare these methods across the entire sample-processing workflow, starting from initial filling of pre-opened SWCNT powders, through controlled removal of excess, unencapsulated molecules, to subsequent dispersion of the material in aqueous media. The resulting suspensions are then subjected to density-gradient ultracentrifugation (DGU) to separate filled from empty nanotubes. Moreover, we demonstrate that leaving the outer surface free enables additional metal–semiconductor sorting via DGU. For comprehensive characterization of the encapsulation efficiency and the resulting electronic modifications, we employ a combination of optical absorption spectroscopy, resonant Raman scattering (RRS), photoluminescence-excitation spectroscopy (PLE), thermogravimetric analysis (TGA), and electron paramagnetic resonance (EPR) spectroscopy. Finally, we evaluate an additional methodology designed to eliminate conventional washing steps by selectively removing unencapsulated molecules without extensive solvent treatment, by vacuum sublimation. This approach aims to minimize interference from externally adsorbed species, thereby enhancing the stability of the encapsulated systems and improving the subsequent sorting steps.

## 2. EXPERIMENTAL SECTION

### 2.1 Materials for filling

#### *2.1.1 Unfilled SWCNT Powders*

For the filling studies, opened SWCNTs are needed, while closed SWCNTs are essential as control references for adsorption on the outside instead of encapsulation inside. To this end, two commercially available SWCNT batches were obtained from Carbon Solutions Inc. A first batch of as-prepared CNTs (AP-SWCNTs, batch #AP-167) consists of a significant fraction of closed SWCNTs and purified CNTs (P2-SWCNTs, batch #03-A019) which were previously found to be completely opened.[31] The degree of opening of this particular batch has been characterized in earlier studies, showing opening rates of more than 90%, as verified by their ratio of closed (empty) and opened (water-filled) SWCNTs when dispersed in an aqueous surfactant solution.[35]

To remove residual functional groups and structural defects that might be introduced during purification, the nanotubes were thermally annealed at 800 °C under vacuum (~$10^{-4}$ Pa). For this vacuum annealing, the temperature was increased gradually following a controlled heating profile (2 h from room temperature to 200 °C, 30 min at 200 °C, 1 h from 200 °C to 800 °C, and 1 h at 800 °C), followed by cooling to room temperature with an initial cooling ramp of approximately 550 °C/h. This treatment yielded undoped, empty, and opened nanotubes[32] ready for the filling procedures. After this treatment, the D/G ratio becomes smaller than 0.015 (peak to peak ratio at 514.5 nm) showing hardly any presence of defects in these annealed and opened SWCNTs (**Figure 2**, D-band at 1350 $cm^{-1}$).

#### *2.1.2 Molecules for filling*

For the filling procedure, commercially available TTF and TCNQ were used. TTF was purchased from TCI Chemicals (Tetrathiafulvalene, purified by sublimation, 99%), and TCNQ was obtained from Fisher Scientific (7,7,8,8-tetracyanoquinodimethane, 98%).

### 2.2 Filling methods

To compare different filling approaches, TTF was used as a model molecule and filling was performed via melting, reflux, and sublimation. The most effective method, vacuum sublimation, was subsequently applied for the encapsulation of TCNQ.

#### *2.2.1 Filling by melting*

A mixture of 20 mg of annealed P2-SWCNTs and at least twice the mass of TTF relative to the SWCNTs was mixed in a mortar to ensure homogenization. The mixture was then transferred to a round-bottom flask equipped with a magnetic stir bar. The system was placed in thermal contact with an oil bath set at ~130 °C, slightly above the melting point (116 - 119 °C) of the dopant. A water-cooled condenser was attached to promote vapor recirculation and prevent material loss, and the entire system was kept under $N_2$ atmosphere, stirring overnight to ensure sufficient time for filling. After cooling to room temperature, a compact pellet was obtained in which the SWCNTs were strongly bundled. Due to the high melting point of TCNQ (~290 °C), melt filling was not feasible for this molecule.

#### *2.2.2 Filling by reflux*

In this approach, 20 mg of annealed P2-SWCNTs were dispersed in 20 mL of a saturated solution of the desired molecule in a suitable solvent (methanol for TTF; solubility of TTF in MeOH was studied prior to the experiment and was found to be around 9.2 g/L at room temperature, see section S1 in the Supporting Materials (SM)). The solution was added to a round-bottom flask, and a magnetic stirring bar was added. The system was purged with nitrogen and maintained under a slight $N_2$ flow before being placed in an oil bath set a few degrees above the solvent boiling point (~65 °C for MeOH). A condenser was used to facilitate continuous solvent recirculation and maintain stable reflux conditions while the solvent was kept at its boiling point. Refluxing under a $N_2$ atmosphere was done overnight to ensure sufficient time for filling. In contrast to melt filling, this method may result in partial solvent encapsulation, which can reduce the effective filling yield. However, reflux remains suitable for molecules that are unstable under melt conditions.

### *2.2.3 Filling by vacuum sublimation*

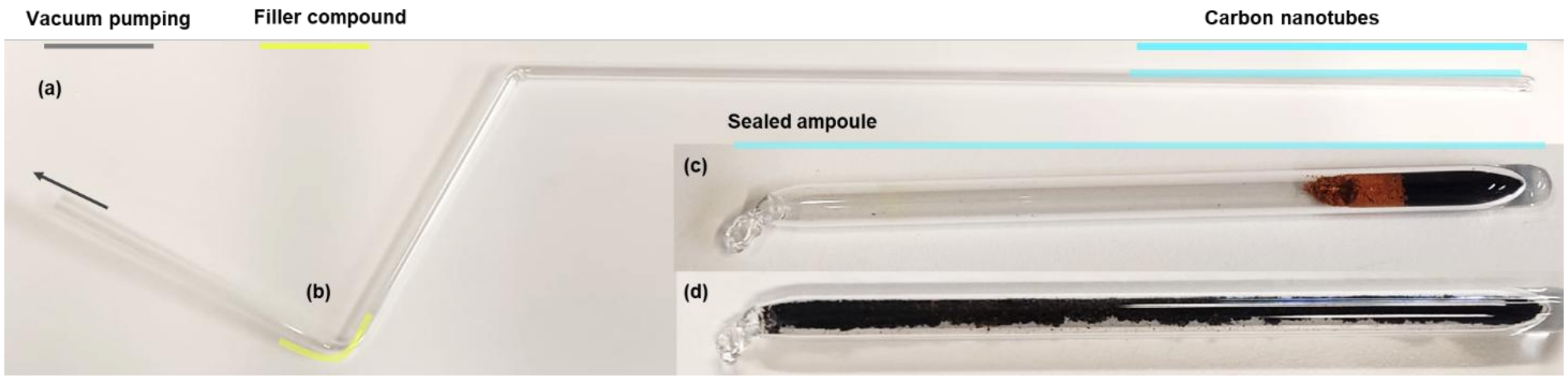

**Figure 1.** (a) Picture of a bent soda-lime glass tube used for filling by sublimation before adding any of the compounds, open on one end to allow degassing of the nanotubes. (b) The filler compound is placed at one corner during the degassing of the nanotubes, which is cooled to prevent evaporation at this stage. (c) The molecule and the carbon nanotubes are then placed together at the other end of the glass tube and sealed under vacuum. The tube is subsequently rotated until a homogeneously mixed powder is obtained (d).

For the sublimation method, 20 mg of annealed P2-SWCNTs were degassed at 300°C for 1 hour under dynamic vacuum in an S-shaped soda-lime glass tube, as shown in **Figure 1**. A dopant mass at least twice that of the nanotubes was introduced into a separate section of the tube and kept at a lower temperature during degassing to prevent premature sublimation. After cooling to room temperature while maintaining the dynamic vacuum, the nanotubes and dopant were brought into the same section of the tube, which was then vacuum-sealed. The sealed ampoule (see **Figure 1d**) was heated for 24 h at a temperature slightly above the sublimation point of the dopant (~100 °C for TTF and ~200 °C for TCNQ), followed by slow cooling inside the furnace to room temperature. The sample ampoule was then opened, and the powder was collected for further characterization and subsequent rinsing of excess molecules.

### *2.2.4 Filling and extraction by sublimation (FES)*

This procedure, inspired by vapor phase infiltration,[42,43] which is derived from the field of atomic layer deposition, is also based on sublimation but allows switching back and forth between dynamic and static vacuum without sealing the ampoule. The setup is shown in **Figure S2** in Supplementary Materials (SM). 40 mg of annealed P2-SWCNTs were degassed at 120–200 °C for 15 min under dynamic vacuum. The vacuum valve was then closed to create static vacuum conditions, and the system was heated above the sublimation temperature of the filler compound for 1–3 h to promote filling. Subsequently, dynamic vacuum was reapplied, maintaining the same temperature used for the filling to remove the excess material on the outer surface of the tubes (1-3h). Different temperatures and times were used, as described in **Table S1** and **Figure S3** in the SM. The temperatures were chosen from TGA analysis of the other samples (**Figure 7**), showing adsorbed species being removed at those temperatures, but optimization of time and temperature should still be conducted in the future. The method relies on the stability of the molecules after encapsulation, thus not removing those, while the externally adsorbed molecules can be removed more easily (see further). The FES method offers the advantage of removing excess molecules directly after the sublimation step without exposing the system to a different atmosphere, while still allowing sufficient time for the filling process to occur before any flushing of residual material. Importantly, this method does not require the use of a solvent to remove the excess of adsorbed molecules remaining on the external nanotube surface, which provides two important advantages, namely, it significantly reduces the sample preparation time (not requiring subsequent extensive rinsing) and moreover, prevents any unwanted solvents from being ingested in the SWCNTs.

***2.2.5 Washing methodology***

For all filling methods except FES, an additional washing step was performed to remove excess molecules from the nanotube surface while minimizing loss of encapsulated material (**Figure S4, SM**). Excess TTF was removed by rinsing and filtration using methanol (MeOH), whereas excess TCNQ was removed using dichloromethane (DCM). For sublimation and melt samples, the solid material was immersed in a known volume of the appropriate solvent, briefly bath-sonicated (~15s, just enough to help the sample redisperse in the solvent, avoiding clumping), and filtered through a water-wettable PTFE (wwPTFE) membrane with a pore size of 0.45 μm, which retains the SWCNTs while allowing excess molecules and solvent to pass through. For reflux samples, the first washing step consisted of directly filtering the reaction mixture. The sample was subsequently removed from the membrane, redispersed in fresh solvent, and refiltered.

Depending on the solubility of the molecules, a significant number of washing steps is needed to fully remove the molecules, however, it was previously shown that also encapsulated molecules can exit the SWCNTs with rinsing.[44] To monitor the washing process and decide when to stop the iterative procedure, the filtrate absorption spectrum was measured after each step to monitor the decreasing concentration of free molecules in the solvent. There is no clear criterion on when to stop the washing, often referred to as 'until the filtrate is colorless' (by eye) or 'until the washing no longer results in a significant amount of molecules being removed'. To obtain a more in-depth monitoring of this washing, in this work, after each washing step, a few milligrams of the sample were separated from the main batch and kept for analysis.

After washing, the samples were dried under $N_2$ atmosphere for three days before solubilization and characterization.

## 2.3 Solubilization and sorting

### *2.3.1 Solubilization*

After washing, the dopant@SWCNT samples were solubilized in an aqueous surfactant solution. A 1% (w/v, 10 mg/mL) sodium deoxycholate (DOC, 99%, Acros Organics) solution was prepared in deuterated water ($D_2O$, 99.8 atom% D, Cortecnet). Subsequently, 20 mg of powdered SWCNTs were added to 4 mL of the surfactant solution. A magnetic stirring bar was introduced, and the dispersion was stirred for at least three weeks. For the first three days of this stirring, bath sonication was applied for 15 min per day to ease the individualization of the SWCNTs.

### *2.3.2 Centrifugation*

Following solubilization, the samples were centrifuged for 22h or 24 h at 16,000 g using a tabletop centrifuge (Sigma 2–16KCH) equipped with a swing-out rotor. This centrifugation is known to remove bundled nanotubes from the dispersion and essential to evaluate the encapsulation in individual SWCNTs so that effects of interstitial molecules between SWCNT bundles can be excluded. The supernatant (approximately 3.5 mL) was carefully collected and used either for subsequent DGU or for characterization.

### *2.3.3 DGU for empty-filled separation*

In some cases (as mentioned in the text), DGU was performed to separate empty and filled CNTs using iohexol (Nycodenz) as the gradient medium, following a previously established procedure.[45] In brief, a step gradient was prepared by layering equal volumes of a low-density iohexol solution (ρ = 1.213 g/mL in $D_2O$) containing the SWCNT supernatant on top of a high-density solution (ρ = 1.289 g/mL in $D_2O$), both containing 1% (w/v) DOC. This was then converted into a continuous gradient by tilting and rolling the centrifuge tube. Ultracentrifugation was carried out for approximately 20 h (sufficient to reach equilibrium) at 236,400 g and 20 °C using a tabletop ultracentrifuge (Optima Max-XP, Beckman Coulter) equipped with an MLS-50 swing-out rotor.

After centrifugation, bands corresponding to empty and filled SWCNTs were observed at different vertical positions according to their different buoyant densities (**Figure S5**). The fractions were carefully extracted using a syringe for optical characterization.

### **2.3.4** *DGU for metallic-semiconducting separation*

A different DGU approach was followed to separate semiconducting from metallic SWCNTs.[46] First, a stepwise density gradient was prepared in a centrifuge tube using a 60% iodixanol stock solution containing 2% SDS in $H_2O$. The gradient consisted of layers containing 40%, 35%, 32.5%, 30% and 27.5% iodixanol, prepared by dilution of the stock solution with $H_2O$ while maintaining the 2% SDS concentration. The resulting gradient resulted in a bottom density of 1.22–1.23 g/mL and a top density of 1.14 g/mL. The SWCNT supernatant after the initial centrifugation step was then carefully layered on top of the gradient. Ultracentrifugation was carried out for approximately 9 h at 50,000rpm (243,000 g) and 22°C using a vertical rotor (P50VT2) in a CP100WX ultracentrifuge (Hitachi Koki Co.).

After centrifugation, bands corresponding to metallic and semiconducting SWCNTs form above and below a mixed fraction, respectively (**Figure S6**). Fractions were collected using a semi-automated system, which allowed sampling in steps of approximately 1.5 mL. Fractions with similar metallic or semiconducting purity were combined and stored for subsequent optical characterization.

#### *2.3.5 Dialysis*

After DGU, samples were dialyzed using either centrifugal dialysis filters or dialysis membranes (10 kDa molecular weight cut-off (MWCO)) to replace the gradient medium with a fresh surfactant solution, enabling proper optical characterization. Moreover, samples in $H_2O$ were exchanged to $D_2O$ at the same time to allow for characterizing the samples over a broader wavelength range.

#### *2.3.6 creating high-density SWCNT films for XPS measurements.*

High density SWCNT – films were produced from DGU-sorted semiconducting fractions of the reference s-SWCNTs, TTF@s-SWCNTs and TCNQ@s-SWCNTs by positive dead-end filtration using a custom-made filtration setup, which has been reported in detail in previous publications.[47,48] Briefly, the setup features an air-tight filtration cell (stainless steel) equipped with a metal sieve and a PTFE cylinder used to confine the filtration volume and hold the membrane in place during the filtration process. It also incorporates microfluidic components, which include regulators, pressure sensors and a Coriolis flow sensor, enabling precise control over the flow across the membrane. Bottled $N_2$ gas is used as a pressurizing gas.

s-SWCNT dispersions in 1% (w/v) DOC were concentrated to give an optical density of approx. 2.5/cm measured at the absorption maximum of the $S_{22}$-transition by ultrafiltration (MilliQ Ultrafiltration cell with BioMax 300kDa). 800 µL of the concentrated dispersions were diluted with deionized water (18.2 MΩ) to a final volume of 4000 µL and were used as starting solution. A hydrophilic coated polycarbonate track-etched membrane (it4ip) with a diameter of 47 mm was used as a filtration membrane. Positioning of the membrane inside the filtration cell, loading of the filtration cell with 4000 µL of starting solution and sealing were performed according to a previously reported step-by-step protocol.[47] A low flow rate of 150 µL $min^{-1}$ was maintained for 3250 µl (21 min) before SWCNTs were pushed down fast using a flow rate of 700 µL $min^{-1}$. Nitrogen flow was continued until the film was dry (see **Figure S8** for the profile of the flow and a picture of the resulting film).

For XPS measurements, these films on polycarbonate membranes were transferred onto 1×1 cm silicon nitride wafers featuring 75 µm square openings. To this end, dry SWCNT films on polycarbonate membranes were first placed in a chloroform bath for 10 min dissolving the polycarbonate membrane and producing a free-floating CNT film. The free-floating film was attached to the silicon nitride wafer by gently lifting it from the chloroform surface. SWCNTs attached to silicon nitride substrates were left in the fume hood in air until residual chloroform was fully removed. A photograph of the resulting films on the silicon nitride substrates can be found in **Figure S9**.

### 2.4 Characterization techniques

### *2.4.1 UV-vis-NIR Absorption spectroscopy*

The spectra were recorded with a Cary 5000 UV–VIS-IR spectrometer in the range of 200-2500 nm. Dispersed samples were measured using a quartz cuvette with a 3 mm optical path length. A baseline correction was applied using the same surfactant solution used to disperse the SWCNTs. An $A/\lambda$ scattering background, with $\lambda$ the wavelength and $A$ a scaling parameter was subtracted from the absorption spectra and the spectra were then normalized on the second optical transition of the SWCNTs.

### *2.4.2 resonant Raman spectroscopy (RRS)*

Samples were measured using a Dilor XY800 triple Raman spectrometer with 1800 gr/mm gratings and using the 514.5 nm laser line from an argon laser (Spectra Physics model 2020). The measurements were taken with a laser spot of approximately 70µm × 2 mm with a sufficiently low power to avoid heating, either 35mW (powders) or 100mW (dispersions using continuous stirring).

### *2.4.3 Photoluminescence excitation spectroscopy (PLE)*

For PLE measurements, the sample was excited using a high-pressure Xe lamp (Edinburgh Instruments, Xe900), used in a pulsed mode (10 ms pulses) to limit integration times to 10ms and thus drastically reduce the effect of dark current with the extended InGaAs photodiode array detector with a sensitivity up to 2.2 µm (Princeton Instruments, OMA V:1024LN-2.2). Excitation wavelengths were selected with a 300 mm focal length monochromator (Acton SpectraPro 2355), while emission spectra were collected through a 150mm focal length spectrograph (Acton SpectraPro 2156) equipped with a 150 g/mm grating. Slits of these spectrometers were chosen to have an average resolution of 6 nm in excitation and 8 nm in emission. Spectra were corrected for instrumental response and concentrations were kept in the correctable (re)absorption regime for both excitation and emission wavelengths (OD < 0.3 in a 3 mm pathlength cell). After measurements, spectra were normalized on their respective peak absorption to directly compare emission quantum yields with each other.

***2.4.4 EPR measurements***

EPR was performed within an Oxford cryostat operated at 2.5 K in a standard X-band spectrometer (Bruker Elexys-II E500, 9.44 GHz). Approximately 1-2 mg of the SWCNTs, dopant@SWCNTs, or dopant powder was added inside a 3 mm quartz tube, and EPR experiments were conducted with a magnetic-field modulation amplitude of 0.1 mT and a microwave power of 2 mW. This microwave power was chosen to avoid microwave saturation of any of the signals. All spectra were recorded with the same integration time and measurement parameters. For the dopant@SWCNT samples, first a broad background from the SWCNTs was subtracted by fitting a polynomial to the background, after which the spectra were normalized on their EPR intensity (by dividing by the (max-min) of the signals).

***2.4.5 TGA measurements***

Thermogravimetric analysis measurements were performed using a Mettler Toledo TGA/DSC 3+ with a ramp of 10°C/min under an air flow of 80 mL/min in an aluminum cup, measuring over a temperature range between 30 °C and 800 or 1000 °C. Measurements were intentionally performed in air such that the relative amounts of the different components (encapsulated and non encapsulated molecules as well as SWCNTs) can be differentiated (as they decompose in different temperature ranges) and quantified.

***2.4.6 XPS measurements***

X-ray photoelectron spectroscopy (XPS) was performed using a Phi Genesis XPS system equipped with an Al Kα X-ray source (1486.6 eV). For each sample, a survey spectrum was acquired over a binding energy range from 0 eV to 1300 eV. In addition, high resolution spectra were measured for the C 1s, N 1s and S 2p region with a step size of 0.1 eV. The measurements were performed on CNT powders right after FES filling or on thin films prepared from semiconducting DGU-sorted SWCNT samples transferred on 1 × 1 cm $Si_3N_4$ substrates, as described in section 2.3.6. The latter was done so that potentially interstitial dopant molecules that could still be present in the bundled CNT powders are removed during the solubilization and DGU-sorting. The Si signal stemming from the $Si_3N_4$ substrate was used as a calibration reference, with the Si 2p binding energy set to 101.7 eV to correct for a potential spectral drift.[49] The X-ray beam spot size was approximately 200 μm. All spectra were acquired using identical measurement parameters. The raw spectra were analysed and fitted using the software CasaXPS. The C 1s spectra were fitted using three asymmetric Lorentzian line shapes, LA (50), together with a Shirley background.

## 3. RESULTS AND DISCUSSION

### 3.1 Washing of externally adsorbed molecules

First, we fill SWCNTs by the vacuum sublimation method with both TTF and TCNQ and compare the SWCNT powders directly after filling and after successive washing steps by RRS. A wavelength of 514.5 nm was chosen as it is in resonance with a series of semiconducting SWCNTs with an average diameter within the P2 diameter distribution. **Figure 2** presents these RRS spectra, zoomed in on the vibrational range of the SWCNT G band, also in resonance with signals from the encapsulated molecules. As a reference, an unfilled yet opened SWCNT sample was used.

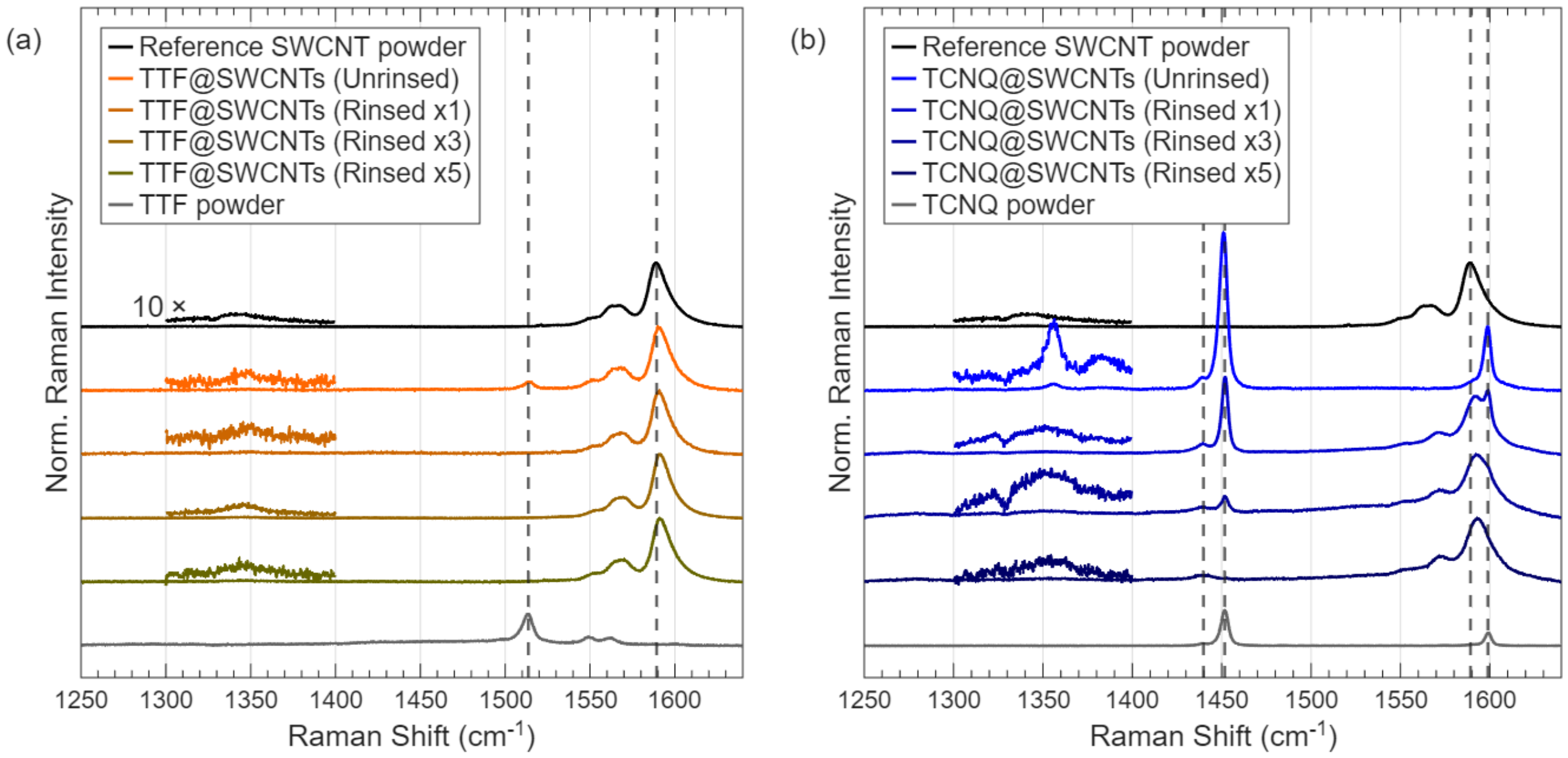

**Figure 2.** (a) Raman spectra at 514.5 nm of TTF@SWCNTs powder sample after no, one, three and five rinsing steps compared to a pristine SWCNTs reference sample and the TTF powder. (b) Raman spectra of TCNQ@SWCNTs powder sample after no, one and five rinsing steps compared to the unfilled, yet opened SWCNT reference sample and the TCNQ powder. Vertically dashed lines are included to visualize any potential shifts of the Raman peaks. The D-band (at 1350 cm$^{-1}$) is enlarged by a factor of 10 as it is barely visible in all these spectra evidencing that not many structural defects are present in the SWCNTs.

In both the TCNQ- and TTF-filled samples, an upshift of the G band is observed relative to the reference sample. This shift persists after several washing cycles. Additionally, Raman features corresponding to the respective molecules are visible in the early rinsing steps. For the TTF-filled sample, these features gradually decrease and disappear after multiple washing rounds, though the G band upshift is retained indicative of remnant doping.[14] For TCNQ, which is more resonant at this excitation wavelength, the unrinsed sample mainly displays peaks from the TCNQ molecules, while the G band of the SWCNTs is hardly visible. After subsequent washing cycles, the peak of the TCNQ at approximately 1440 $cm^{-1}$ remains visible, while the originally much stronger peaks at 1450 $cm^{-1}$ and ~1600 $cm^{-1}$ are fully removed and the G band of the SWCNTs becomes dominant. We tentatively attribute the 1440 $cm^{-1}$ feature to encapsulated TCNQ aggregates, potentially exhibiting J-aggregate-like packing with the confined nanotube cavity.[34] Importantly, even though the TCNQ on the outside seems largely removed (see further), the G band upshift due to doping is still present after extensive rinsing.

The upshift of the G band could arise from several factors, including charge transfer (doping), strain, or laser-induced heating. Prior to the measurements, the G band position was monitored as a function of laser power, and sufficiently low laser powers were used during measurements to exclude effects of laser-induced heating. To disentangle strain-induced shifts from doping-induced shifts, we monitored both the G band and the 2D Raman band (**Figure 3**), which are known to both upshift with doping, but typically behave oppositely with strain, as first demonstrated for graphene[50] and later for SWCNTs.[41] Electrochemical doping studies have shown that the G band upshift is expected for larger-diameter SWCNTs for both p- and n-type doping.[14] Therefore, the observed persistent upshift of both the G and 2D band is consistent with charge-transfer doping of the nanotubes, and not with a modification of the strain.

### 3.2 Comparison of different filling methods

Since TTF filling can be achieved by the three different filling methods, being a molecule with a good solubility in MeOH and sufficiently stable for filling from the melt and by vacuum sublimation, we compare the different filling methods for TTF. **Figure 3** presents the Raman G band and 2D band of TTF-filled SWCNT samples prepared using different filling methods after being extensively rinsed according to the above-mentioned procedure. The experimental data are fitted with a sum of Lorentzians to extract peak positions, which are tabulated in **Table 1**. An upshift of both the G and 2D bands is observed for all filling methods. However, the magnitude of the shifts varies depending on the filling approach. The sublimation method produces the largest overall G band upshift, while the reflux and melt method results in the smallest G shift. This can be easily understood as in the refluxing method, also the solvent can be encapsulated, preventing the SWCNTs from being compactly filled with the dopant molecules or at least slowing down the filling process. While the melt-produced samples display similar 2D band shift, their G band is less shifted. The melt-procedure results in a compact pellet of SWCNTs, difficult to rinse and disperse, most likely resulting in a larger amount of strongly bundled SWCNTs and therefore producing a different G band shift. Since we subsequently want to isolate the SWCNTs in a suspension for further separation purposes, these results show that the vacuum-based filling method seems most efficient in filling the SWCNTs with the dopant molecules while also allowing for further processing.

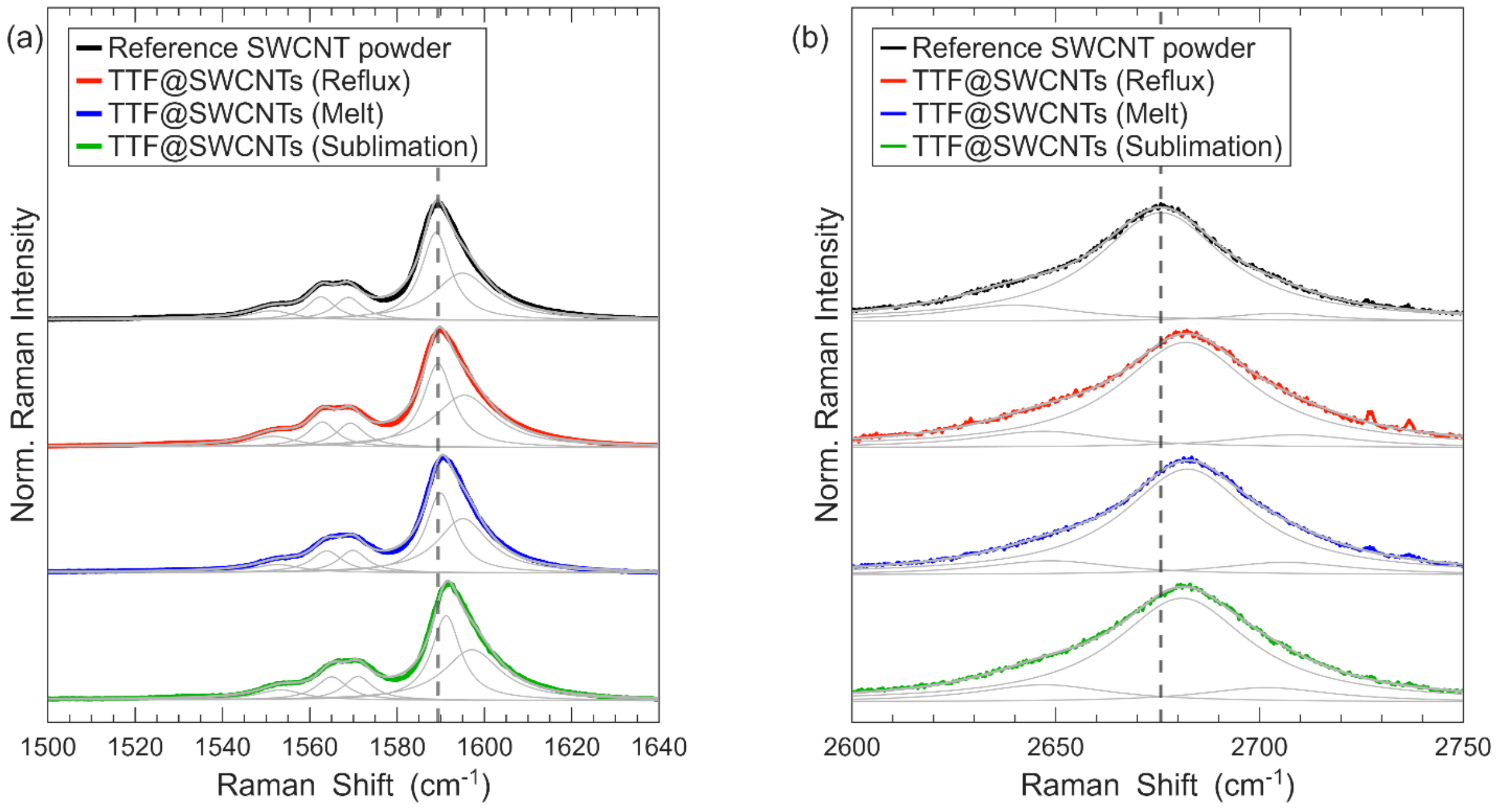

**Figure 3.** Raman spectra at 514.5 nm, showing the upshift of G and 2D bands for all filling methods compared to a reference SWCNT sample. Spectra are individually normalized and vertically shifted for clarity. Both spectra were fitted with a sum of Lorentzians. Note that in bundled, multi-chirality samples typically a superposition of multiple lines for the G and 2D bands are observed,[41] of which the position of the main G and 2D peak is given in Table 1.

**Table 1**: Fitted peak positions of the main G and 2D bands of the Raman spectra of TTF@SWCNT powders from reflux, melt and sublimation-based filling.

| Sample | G band ($cm^{-1}$) | Shift G band ($cm^{-1}$) | 2D band ($cm^{-1}$) | Shift 2D band ($cm^{-1}$) |
|---|---|---|---|---|
| Reference SWCNT powder | 1588.94 ± 0.04 | - | 2675.9 ± 0.1 | - |
| TTF@SWCNT (reflux) | 1589.38 ± 0.04 | 0.44 | 2682.0 ± 0.2 | 6.1 |
| TTF@SWCNT (melt) | 1589.94 ± 0.05 | 1.0 | 2682.3 ± 0.1 | 6.4 |
| TTF@SWCNT (sublimation) | 1591.24 ± 0.04 | 2.3 | 2681.0 ± 0.3 | 5.1 |

However, powder samples are inherently inhomogeneous, and the position of the laser spot on the sample during Raman measurements may introduce local variations in the results. To properly benchmark the different filling methods and obtain statistically more robust data, the powders were dispersed in a surfactant solution to isolate the SWCNTs individually. This approach creates a more homogeneous environment for the SWCNTs and minimizes local effects, enabling a more reliable comparison between filling methods. To prepare the samples for absorption measurements, thoroughly rinsed TTF- and TCNQ-filled SWCNT powders were dispersed in an aqueous surfactant solution of 1% wt/V DOC at the same mass-to-volume ratio (10 mg/mL). After 24 h of centrifugation to remove most bundles, the dispersions were analyzed to evaluate both the quality of the dispersion and the effect of filling on the electronic structure of the nanotubes.

Clear differences in dispersion were observed. The sample filled with TTF via the melting method exhibited significantly poorer dispersion compared to those prepared via sublimation or solvent reflux, resulting in a lower optical density (**Figure S7**). This behavior is most likely related to recrystallization of the material during the melting and subsequent cooling process. Compact recrystallized aggregates can form dense pellets that are difficult to redisperse and challenging to rinse thoroughly. Experimentally, the recrystallized material obtained after melting appears more resistant to washing, suggesting that residual external TTF crystals may remain in the powder. These residual crystals could negatively affect both dispersion efficiency and optical measurements.

### 3.3 Characterizing the doping of SWCNTs by optical spectroscopy

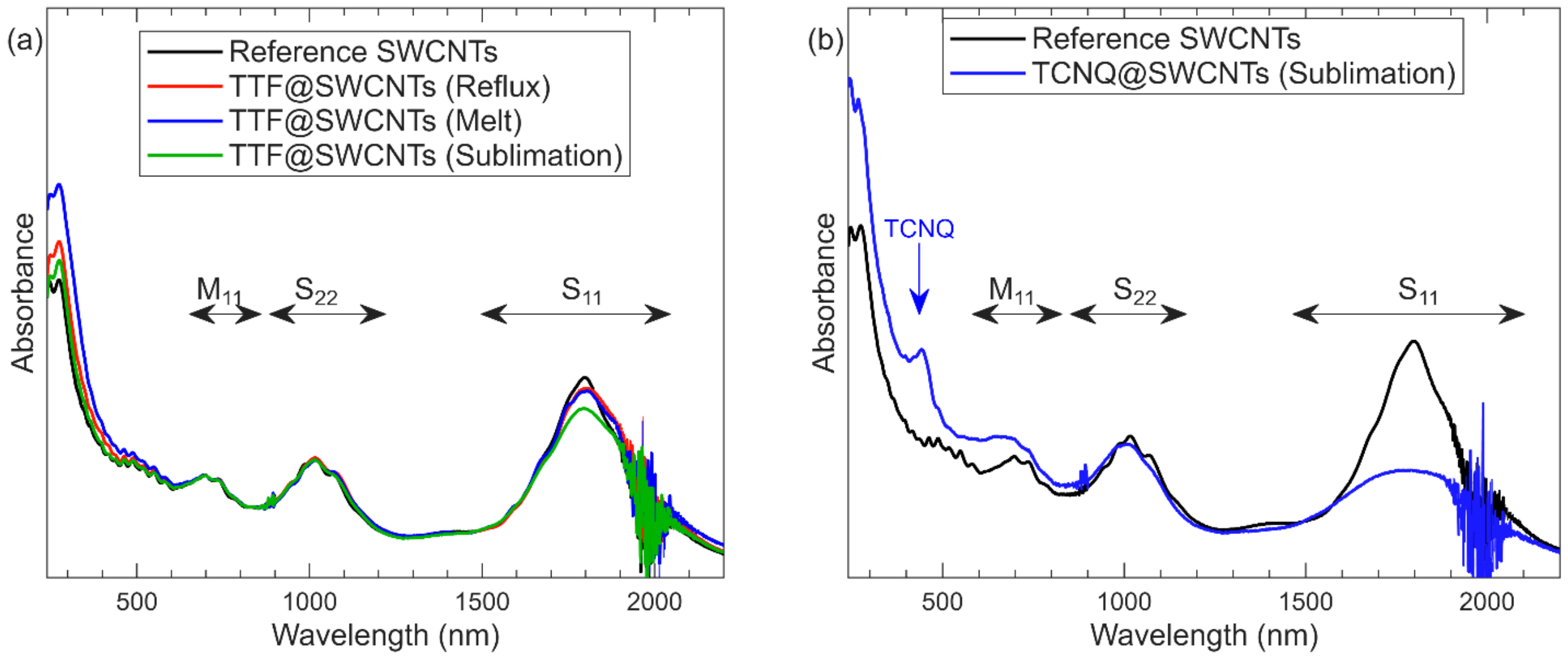

**Figure 4.** Absorption spectra of the dispersed dopant@SWCNTs samples compared with a reference SWCNT sample, after a 1/wavelength baseline subtraction and normalization on the $S_{22}$ absorption peak (around 1000 nm). (a) absorption spectra for the different TTF@SWCNT samples and (b) absorption spectra for the TCNQ@SWCNTs sample. Ranges for the first ($S_{11}$) and second ($S_{22}$) transitions of the semiconducting SWCNTs and the first of the metallic ($M_{11}$) are also indicated.

While in the powder samples doping could still be induced by molecules encapsulated in the interstitial space in SWCNT bundles, we dispersed the SWCNTs in an aqueous surfactant solution and performed centrifugation to remove any bundles and only keep individually dispersed SWCNTs in solution. Figure 4 provides a direct comparison of the absorption spectra of the dispersed filled samples with a reference sample consisting of identically opened SWCNTs dispersed under the same conditions (hence, mostly water-filled). Variations in the $S_{22}/S_{11}$ intensity ratio can be used to monitor shifts in the Fermi level. In particular, bleaching of the $S_{11}$ transition relative to $S_{22}$ is indicative of doping.[21] Since bleaching occurs symmetrically for both n-type and p-type doping, complementary techniques are required to unambiguously determine the doping direction. In this study, however, the known donor or acceptor character of the encapsulated molecules used for filling allows the doping type to be reasonably inferred, as well as confirmation from EPR experiments (see further). Analysis of the $S_{22}/S_{11}$ intensity ratio therefore, provides a practical approach to qualitatively evaluate charge transfer and hence compare the relative filling efficiency of the different methods.

After normalizing the spectra to the $S_{22}$ transition of the CNTs, bleaching of the $S_{11}$ transition is observed for all TTF- and TCNQ-filled samples relative to the reference. For TTF, the bleaching is much smaller compared to TCNQ. However, this would be expected, because TTF first needs to compensate for the external p-type doping of the SWCNTs (in aqueous environment) after which n-type doping can be achieved. For TTF, the bleaching varies in intensity depending on the filling method, with the vacuum sublimation method exhibiting the strongest effect. The observed bleaching trends are consistent with the Raman observations in the powders prior to solubilization, further supporting that the sublimation method induces the highest degree of doping and, consequently, the most efficient filling. Regarding the differences between the melt and reflux methods, the smaller doping effect observed for the reflux method is consistent with partial or mixed solvent filling. Additionally, the practical difficulties associated with washing and dispersing melt-filled powders highlight the advantages of the sublimation method, which provides both more efficient filling and better dispersibility.

For the TCNQ-filled SWCNT sample, two additional absorption features can be observed: a peak around 445 nm and a broader feature around 620 nm, overlapping with the metallic-transitions of the SWCNTs. **Figure 5** provides a comparison between several reference samples and the TCNQ-filled SWCNT sample. From the subtraction of the TCNQ-filled SWCNT sample and the reference SWCNT sample, it can be clearly observed that the peak around 445 nm is red-shifted with respect to the TCNQ in the reference samples (400 nm) whether TCNQ is dispersed in dichloromethane (DCM) or is mixed with closed SWCNTs in an aqueous surfactant solution. Hence, this strong red-shift is a clear indication of encapsulated TCNQ molecules. Such red-shifts have also been observed for other dye molecules after the formation of J-aggregates inside SWCNTs.[34] On the other hand, the broader feature around 640 nm is also visible in the surfactant-solubilized TCNQ reference sample and remains unshifted, thus most likely originates from free TCNQ molecules that are not interacting with the SWCNTs. Literature studies have shown that dimerization of TCNQ can occur in surfactant solutions, resulting in this broadened, red-shifted absorption peak.[51] These results thus seem to point out that not all externally adsorbed TCNQ molecules have been removed by the extensive rinsing and that surfactant-stabilized TCNQ aggregates are still present in the sublimation samples (see also **Figure 9,** which will demonstrate that if there are no external molecules present, this band is not observed).

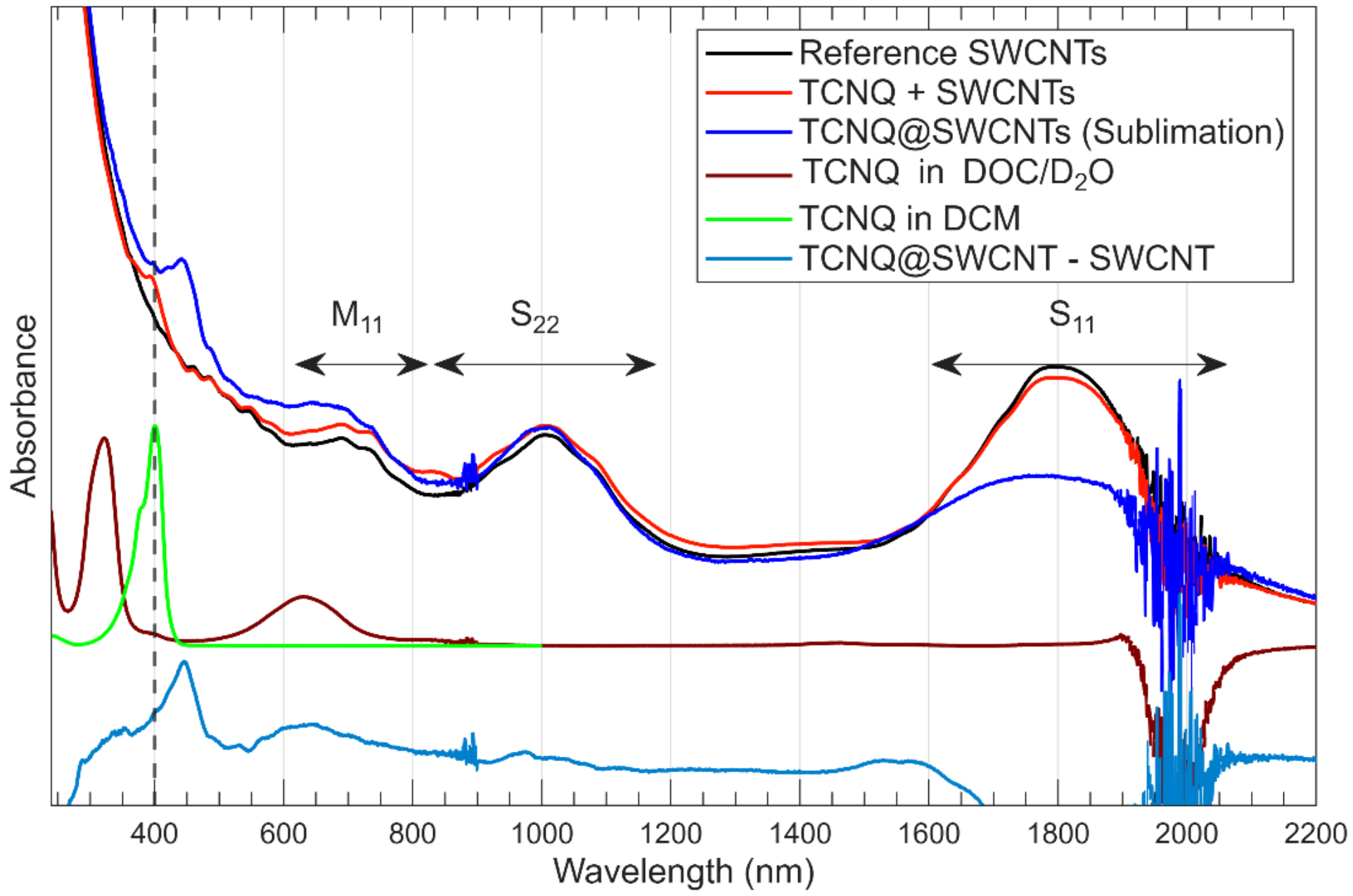

**Figure 5.** Absorption spectra of TCNQ (reference) samples: reference closed SWCNTs, closed SWCNTs mixed with TCNQ, TCNQ@SWCNTS (Sublimation), and free TCNQ in 1% DOC/$D_2O$ and in DCM. Moreover, the subtraction of TCNQ@SWCNT minus the reference SWCNTs is also shown to plot the spectrum of the encapsulated molecules.

The presence of this feature suggests that the powders used to prepare the dispersions still contain residual TCNQ, indicating that the washing procedure did not fully remove externally adsorbed molecules. Another possibility is minor leakage of TCNQ from the SWCNTs during dispersion; however, the low solubility of TCNQ in water and most other solvents makes this scenario less likely. As we will show later, TGA analysis shows the presence of remnant dopant molecules on the outside of the SWCNTs.

The extinction coefficient of TCNQ in DCM at 400 nm was experimentally determined to be 56,500 $Lmol^{-1}cm^{-1}$ (**Figure S1a**). For comparison, encapsulation studies of the DANS molecule report an extinction coefficient of 28,500 $Lmol^{-1}cm^{-1}$ along with a redshift of the absorption peak upon encapsulation.[33] The higher extinction coefficient of TCNQ relative to DANS explains that even small amounts of encapsulated TCNQ can produce detectable absorption features, supporting the assignment of the 446 nm peak to molecules inside the CNTs.

TTF has a much lower extinction coefficient (13,600 $Lmol^{-1}cm^{-1}$ at 300 nm) and its absorption lies in the UV region (see **Figure S1a**), which limits the applicability of the same analysis used for TCNQ on the TTF-filled samples. However, the observed shifts in spectral features of the SWCNTs in the TTF-filled samples (see **Figure 6 and Figure 11)** are consistent with filling, as also supported by Raman spectroscopy and TGA measurements.

Raman measurements were conducted on the same set of samples after dispersion. **Figure 6** shows the Raman spectra at 514.5 nm in the RBM region. Shifts in the RBM modes provide additional confirmation of filling. Because water filling also causes RBM upshifts, a water-filled reference was used alongside an empty/closed-SWCNT reference.[45] The RBM shifts observed in the TTF-filled samples are distinct from those caused by water filling, indicative of filling with another molecule.[35] The RBMs of the TCNQ-filled samples were not measured due to strong fluorescence from residual free TCNQ in solution at this excitation wavelength, which prevented reliable spectral acquisition.

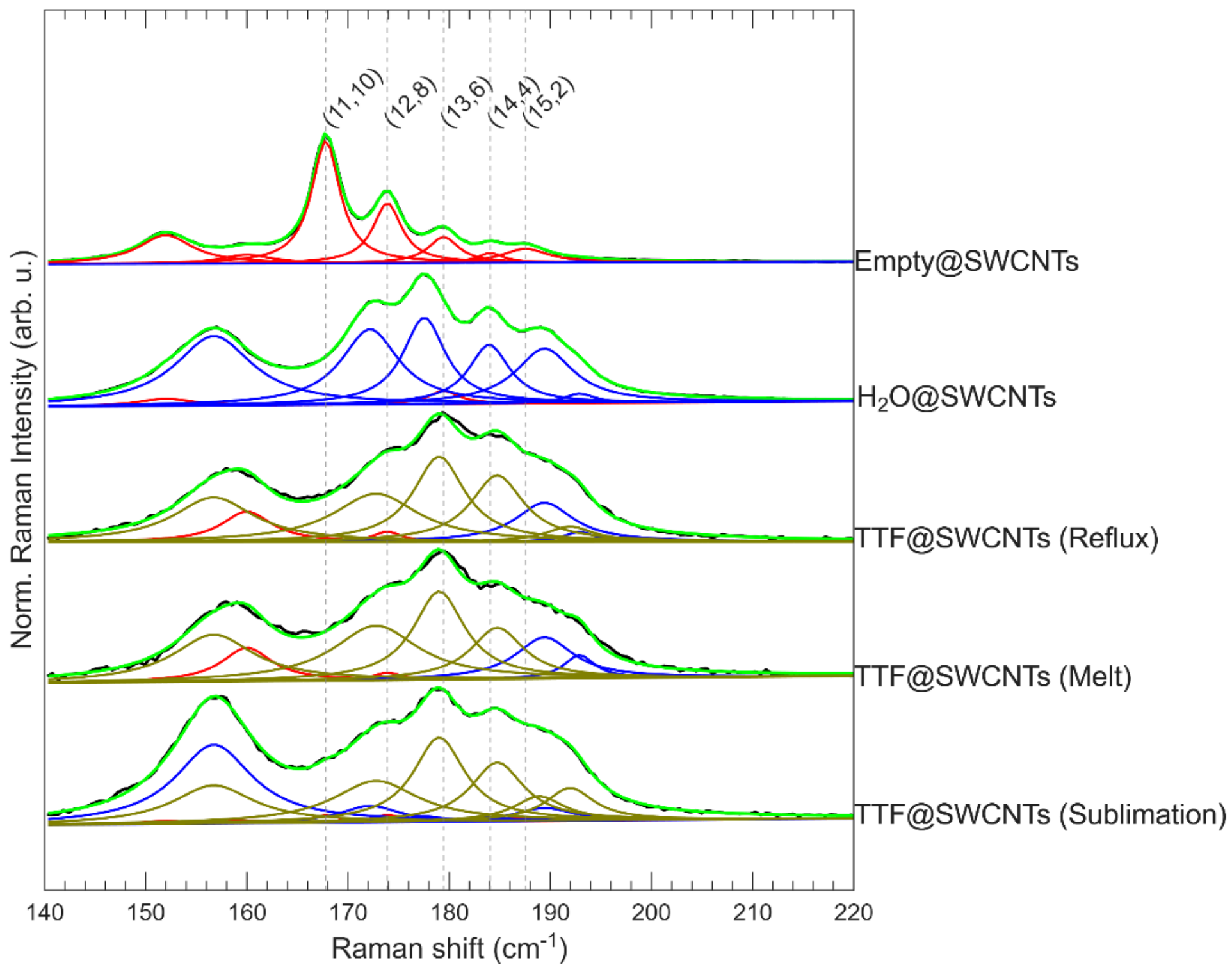

**Figure 6.** Raman spectra of the RBM region at 514.5nm of empty, water-filled and TTF-filled surfactant-solubilized samples with superimposed fits consisting of a superposition of empty (red), water-filled (blue) and TTF-filled (brown) peaks.

### 3.4. Assessing the filling yield by TGA

TGA measurements were performed to quantify both the encapsulated and externally adsorbed material in the filled SWCNT samples. **Figure 7** shows the TGA curves of the reference SWCNTs as well as the TTF- and TCNQ-filled samples prepared by the sublimation method. To clearly identify the temperatures of the mass-loss events, the derivative of each TGA curve was computed from which the maxima can be derived by fitting.

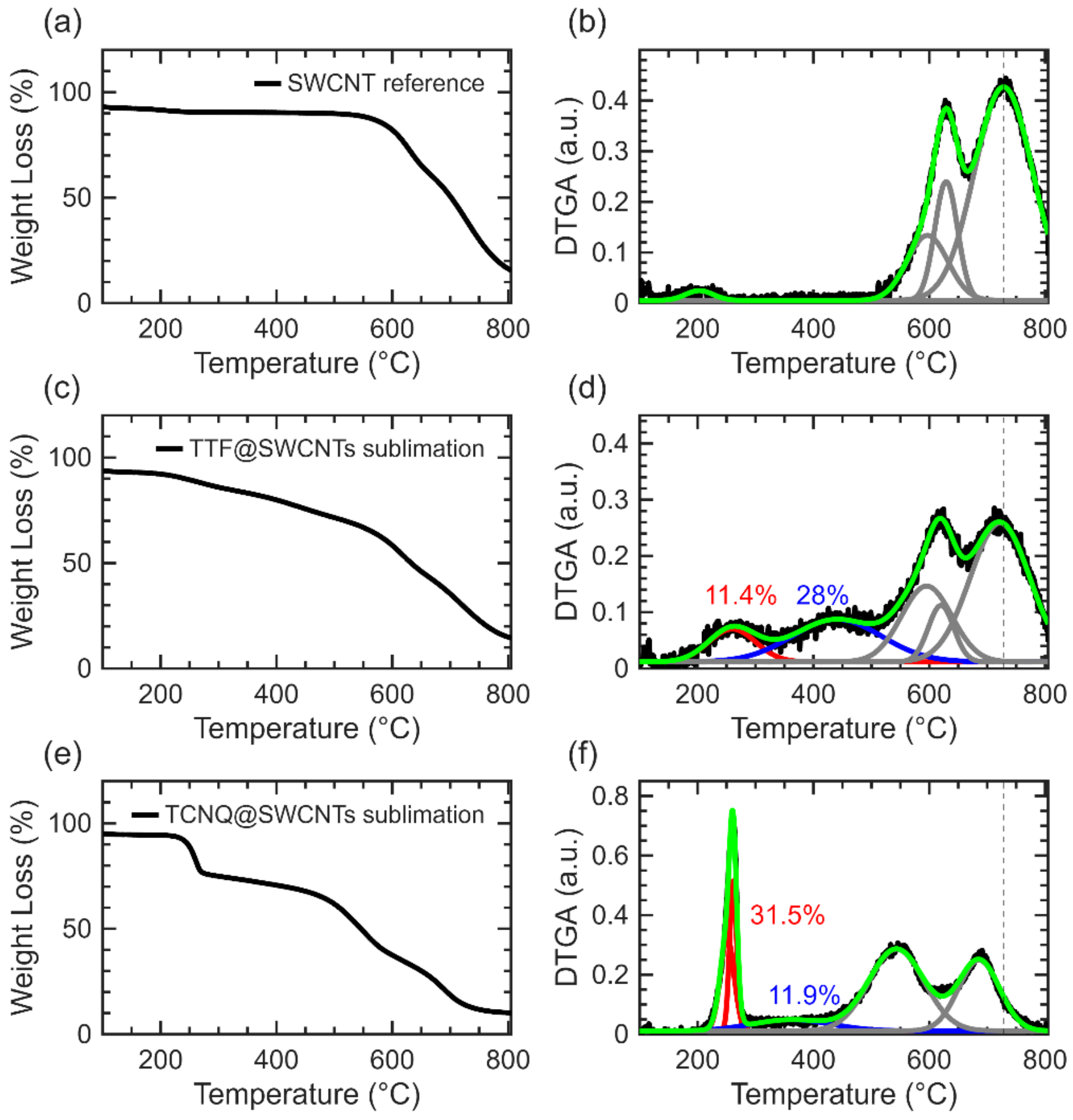

**Figure 7. TGA Analysis of sublimation samples** (a), (c) and (e) show the measured TGA curves for the different samples while (b), (d) and (f) show the derivative of the TGA curves fitted with a sum of Gaussian functions (green; separate Gaussian components are shown in red, blue and grey). (a-b) Reference unfilled, yet opened SWCNT without dopant, (c-d) TTF-filled via sublimation and extensive rinsing, (e-f) TCNQ-filled via sublimation and extensive rinsing. The relative masses of the non-encapsulated (red) and encapsulated (blue) dopant molecules are also provided, calculated as $M_{dopant}/M_{SWCNT}$, thus excluding the initial mass loss below 100˚C and the mass of the metal catalyst nanoparticles. The vertical dashed line indicates the highest combustion temperature observed for the reference SWCNTs.

The reference, pristine SWCNTs exhibit two major mass losses at 629°C and 728°C, corresponding to the thermal decomposition of the nanotube material, and a minor mass loss at 200°C, associated with surface-bound or residual species. Moreover, a non-combustable residual mass of about 15.5% was observed originating from the metallic nanoparticles used in the synthesis. The most noticeable differences between the reference and the filled samples are the appearance of additional mass-loss features, at lower temperatures. For both dopants, two additional peaks are observed, one at a lower temperature that can be attributed to the externally adsorbed dopants (visualized in red in **Figure 7**) and a second peak at a slightly higher temperature, which we assign to encapsulated molecules (visualized in blue in **Figure 7**). The temperature difference between the peaks is consistent with the chemical protection provided by the CNTs to the encapsulated molecules. Gaussian fitting of the derivative TGA curves allows for the estimation of the relative mass percentages of dopants relative to the SWCNT mass (excluding initial mass losses below 100°C and the residual mass from nanoparticles in the SWCNT material, thus $M_{dopant}/M_{SWCNT}$). Using this approach, for the TCNQ-filled sample the encapsulated material mass was approximately 11.9 % of the SWCNT mass. In contrast, the TTF-filled samples prepared via sublimation exhibited higher filling, with an encapsulated material mass estimated to be 28 % of the SWCNT mass.

To estimate the expected filling weight percentage for this diameter range, we chose the (14,6) chirality as a typical diameter within the diameter range of the P2 SWCNTs. A (14,6) SWCNT has a diameter $d$ = 1.3917 nm and a unit cell $t$ = 5.538 nm, which contains 1352 carbon atoms. The mass of a (14,6) SWCNT per unit length can thus be calculated to be $1.6314 \cdot 10^{-21}$ g/nm. For a single file of TTF molecules with a molar mass of 204.36 g/mol, the length of a TTF molecule, including standard van der Waals radii, is around 1.105 nm, making the total mass per unit length for a closely packed single-file array $3.0711 \cdot 10^{-22}$ g/nm In other words, for a close packed filling with a single file of molecules, an $M_{TTF}/M_{SWCNT}$ of 9% is expected, while for a double file of molecules, an $M_{TTF}/M_{SWCNT}$ of 18.5% is expected. Although this is a simple calculation, assuming a single diameter for the diameter distribution in the sample, this value corresponds nicely to the observed filling yield for TTF, hence corresponding to a complete filling with a double row of TTF molecules. In contrast, for TCNQ it seems that rather a single-file filling is achieved.

On the other hand, the TCNQ-filled sample still contains a large fraction of non-encapsulated molecules (31.5%), while for TTF a lower percentage is observed (11.4%). This can be attributed to the lower solubility of TCNQ (2.45 mg/L) in the rinsing solvent compared to TTF (9.2 g/L), resulting in less efficient washing. The TGA experiments do indicate that these molecules can be easily removed by oxidation of the filled samples in air to intermediate temperatures. **Figure 8** presents TGA measurements obtained after 1h of air oxidation (at 220°C for TTF and 260°C for TCNQ). The observed mass-loss percentages indicate that nearly all externally adsorbed material can be removed without harming too much the encapsulated species. This methodology therefore provides a practical approach to selectively remove surface-bound dopants while preserving encapsulated species.

Aside from the additional peaks for the encapsulated and non-encapsulated dopants, the TGA experiments also demonstrate an effect on the SWCNT stability upon doping and oxidation. In **Figure 7** it can be clearly seen that for the TCNQ samples, the SWCNT peaks are shifted to significantly lower temperatures. Moreover, upon oxidation to remove the externally adsorbed dopant molecules (**Figure 8**), the SWCNT combustion peaks change significantly, where the lower combustion peak becomes stronger in amplitude than the second combustion peak. This leaves us to conclude that this oxidation, even when performed at much lower temperatures than the combustion peaks of the SWCNTs, damages the SWCNT structure, which is not beneficial if future nanoelectronics applications are aimed for. We therefore sought a different method to remove these externally adsorbed molecules, which was found in the filling and extraction by sublimation method,  referred to as FES.

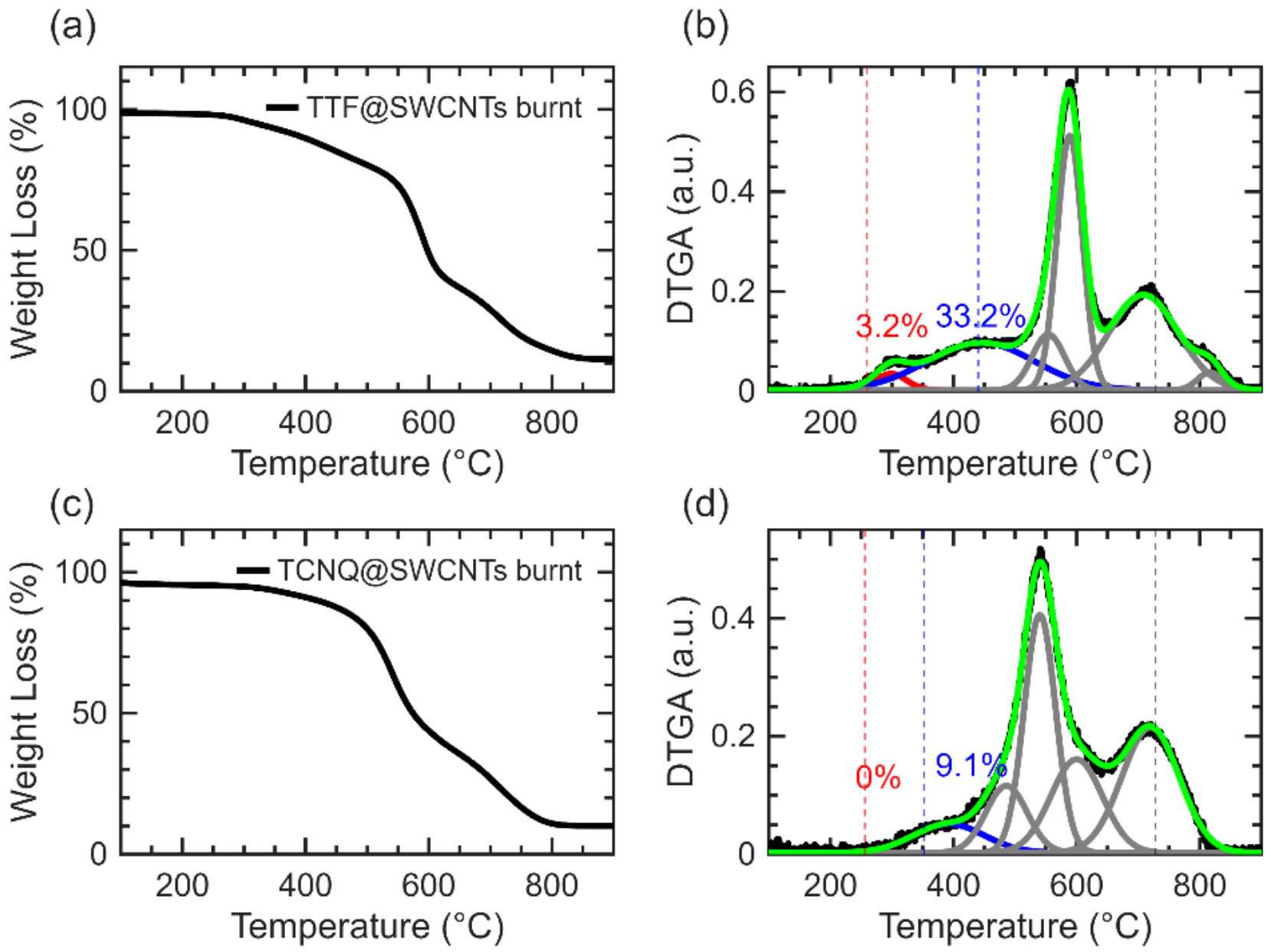

**Figure 8. TGA analysis of oxidized, sublimation-filled SWCNTs**. (a,c) measured TGA curves and (b,d) derivative TGA curves, fitted with a sum of Gaussian functions (green; separate Gaussian components are shown in red, blue, and grey). (a-b) TTF-filled SWCNT samples via sublimation after extensive rinsing and subsequent oxidation at 220˚C for 1h. (c-d) TCNQ-filled SWCNT samples via sublimation after extensive rinsing and subsequent oxidation at 260˚C for 1h. The relative masses for the non-encapsulated (red) and encapsulated (blue) dopant molecules are also provided, calculated as $M_{dopant}/M_{SWCNT}$, thus excluding the initial mass loss below 100˚C and the mass of the metal catalyst nanoparticles. The vertical dashed line indicates the highest combustion temperature observed for the reference SWCNTs, while the red and blue dashed lines indicate the positions of the two dopant peaks before oxidation (**Figure 7**).

### 3.5 FES for external dopant removal

FES consists of a first sublimation filling step, similar as performed above, after which the static vacuum is changed to a dynamic vacuum through which the external molecules can be removed. Different parameters can be optimized, such as the filling temperature, the removal temperature, and the time of the filling and removal cycles. For TTF, we found that a filling temperature of 250°C was more efficient than a filling temperature of 200°C, and that a filling step of 1 hour and a removal step of 3 hours was sufficient to obtain encapsulated species (**Figure S2**). For TCNQ, a similar approach was followed. **Figure 9a** and **9d** present the DTGA curves with superimposed Gaussian fits for both samples. Interestingly, hardly any non-encapsulated molecules are found in this manner for both TTF and TCNQ filled samples (4% for TTF and a non-detectable amount for TCNQ), while still reaching a satisfactory filling yield for both dopants (17% for TTF and 9% for TCNQ; compared to 28% and 11.9% by sublimation filling and rinsing but the latter with excess outside molecules). Corresponding absorption spectra are provided in **Figure 9b,e** and show the expected bleaching of the $S_{11}$ peak with respect to the $S_{22}$ peak, to a similar extent as for the sublimation samples, though now the additional broad absorption band from the non-encapsulated TCNQ species is no longer present. Finally, the G band indicates similar shifts as for the sublimation samples (**Figure 9c,f**), in agreement with intentional doping of the SWCNTs. Thus, vacuum sublimation followed by a removal of the externally adsorbed molecules under dynamic vacuum seems to be the best method to maintain sufficiently high filling yields while largely removing externally adsorbed molecules. A complete parametric optimization of the various parameters in the FES method is however, kept for a follow-up study.

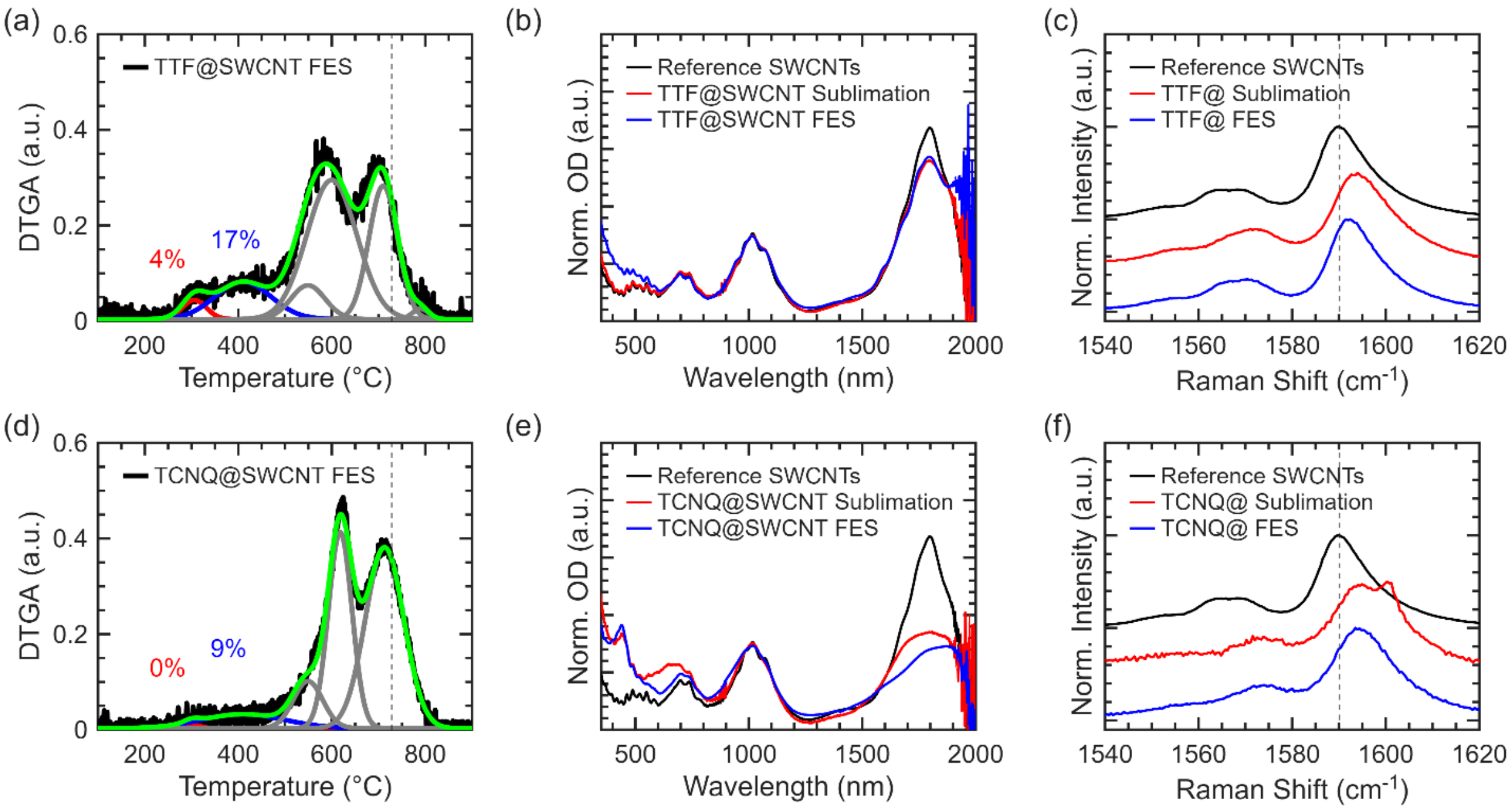

**Figure 9** (a,d) DTGA curves (black) and Gaussian fits (green) of the different contributions of encapsulated (blue) and non-encapsulated (red) dopant molecules and SWCNTs (grey) for (a) TTF@SWCNT prepared by FES and (d) TCNQ@SWCNT prepared by FES. (b,e) Absorption spectra of the surfactant solutions after 22h of centrifugation for reference SWCNTs, sublimation and FES prepared SWCNTs. (c,f) Raman spectra at 514.5 nm of the powders of reference SWCNTs, sublimation and FES prepared SWCNT samples.

### 3.6 Evaluation of p- and n-type doping through EPR spectroscopy

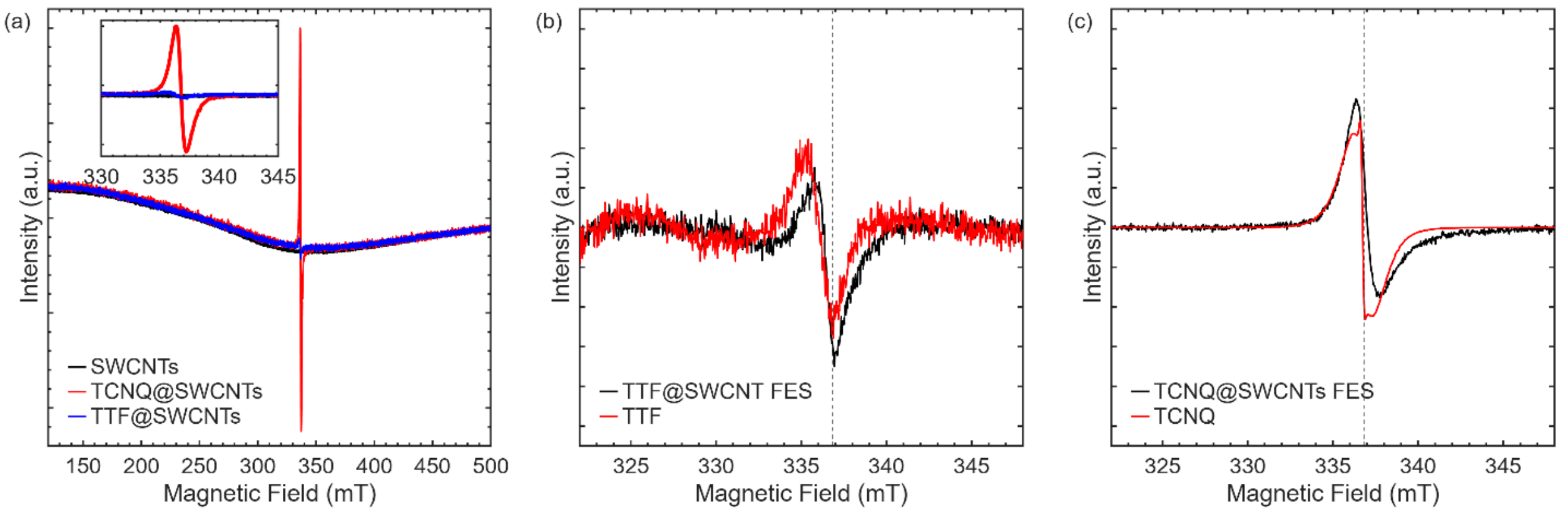

**Figure 10.** EPR spectroscopy (a) Full field range EPR spectra of the reference undoped SWCNTs (black), TCNQ-filled (red) and TTF-filled SWCNTs (blue) with the inset zoomed in on the $g = 2$ region. Spectra are normalized on the broad SWCNT background. (b) Comparison of TTF@SWCNTs through FES with oxidized TTF. The grey line indicates the free-electron $g$-value. Spectra were normalized for comparison. (c) Comparison of TCNQ@SWCNTs through FES with reduced TCNQ. The grey line indicates the free-electron $g$-value. Spectra were normalized for comparison.

**Figure 10a** compares the EPR spectra of the FES-doped SWCNT samples with those of undoped SWCNTs used as a reference (similar results were obtained for the gas-phase method). The reference material exhibits a broad background signal, which is consistent with residual catalytic nanoparticles typically present in SWCNT powders.[44,52] In contrast, the doped samples show a distinct narrow resonance near the free-electron $g$-value. When the spectra are normalized to the broad SWCNT background, this signal is significantly more pronounced for the TCNQ-doped sample than for the TTF-doped one, in agreement with more significant doping observed for TCNQ in optical spectroscopy.

A closer examination of this narrow feature reveals contributions from oxidized TTF species (**Figure 10b**), with $g = 2.0066$ for oxidized TTF and $g = 2.0052$ for TTF@SWCNT, thus slightly higher than that of the free electron. On the other hand, a $g = 2.0023$ was found for reduced TCNQ and a $g = 2.0009$ for TCNQ@SWCNT (**Figure 10c**), thus near the free-electron $g$-value. Control samples of TTF and TCNQ, prepared by annealing under conditions identical to those used during the filling procedure but without SWCNTs, serve as reference samples. Moreover, the higher $g$-value for positively charged TTF molecules is in agreement with expectations for positive versus negative charge carriers in EPR in organic semiconductors. Notably, these molecular signatures are detectable only up to approximately 50 K, and no additional signals associated with the SWCNTs themselves are observed. This contrasts with earlier studies in which external $AuCl_3$ adsorption produced complementary EPR features on the SWCNTs, due to localisation of the charge carriers on the SWCNTs, which might not be the case in the endohedrally doped SWCNTs.[22] Nonetheless, the presence of the oxidized TTF and reduced TCNQ resonances clearly demonstrates that the encapsulated molecules undergo charge transfer after incorporation into the nanotubes.

### 3.7 Metal-semiconductor sorting of filled and doped SWCNTs

A key parameter for future devices is the ability to subsequently sort the filled and doped SWCNTs by their electronic character. While semiconducting SWCNTs are required as the active conduction channels in devices such as transistors, metallic SWCNTs introduce unwanted leakage currents and degrade switching performance. Therefore, efficient separation of metallic and semiconducting species remains essential for the fabrication of high-performance SWCNT-based electronics. Since the dopant molecules are embedded inside the SWCNTs and separations are mostly relying on the interaction of surfactants with the outer SWCNT surface, separation protocols can be easily transferred to the doped SWCNTs. To evidence this, we performed density gradient ultracentrifugation to separate metallic and semiconducting doped SWCNTs from each other, using a previously established procedure (see **Figure S5** in the SM for a photograph of the sorted fractions).[46] For TTF-filled SWCNTs a very similar purity is obtained as for the reference SWCNTs, while the TCNQ-filled samples display a slightly lower purity. This can most likely be solved by either doing an additional iterative DGU separation, or by adjusting the fraction selection for the TCNQ-filled samples. Absorption spectra of the separated semiconducting samples are provided in **Figure 11a**, and **Figures 11c-e** provide their photoluminescence-excitation (PLE) spectra. Interestingly, the emission is quasi fully quenched for the TCNQ-doped SWCNT samples, while the TTF-filled SWCNT samples display a very similar PL quantum yield as the reference P2-samples, though with slightly shifted emission peaks with narrower line widths (**Figure 11b**). The quenching is a consequence of Auger recombination of the excitons, transferring their energy to the free charge carriers and thereby recombining non-radiatively.[53] The spectral shifts indicate that the TTF-filled SWCNTs are indeed filled with something other than water, the latter being the filler of the reference SWCNTs, even though their photobleaching is no longer clearly observed due to the counteracting effect of the surfactant solution and the fact that water-filling also already causes a significant reduction in fluorescence quantum efficiency.[35]

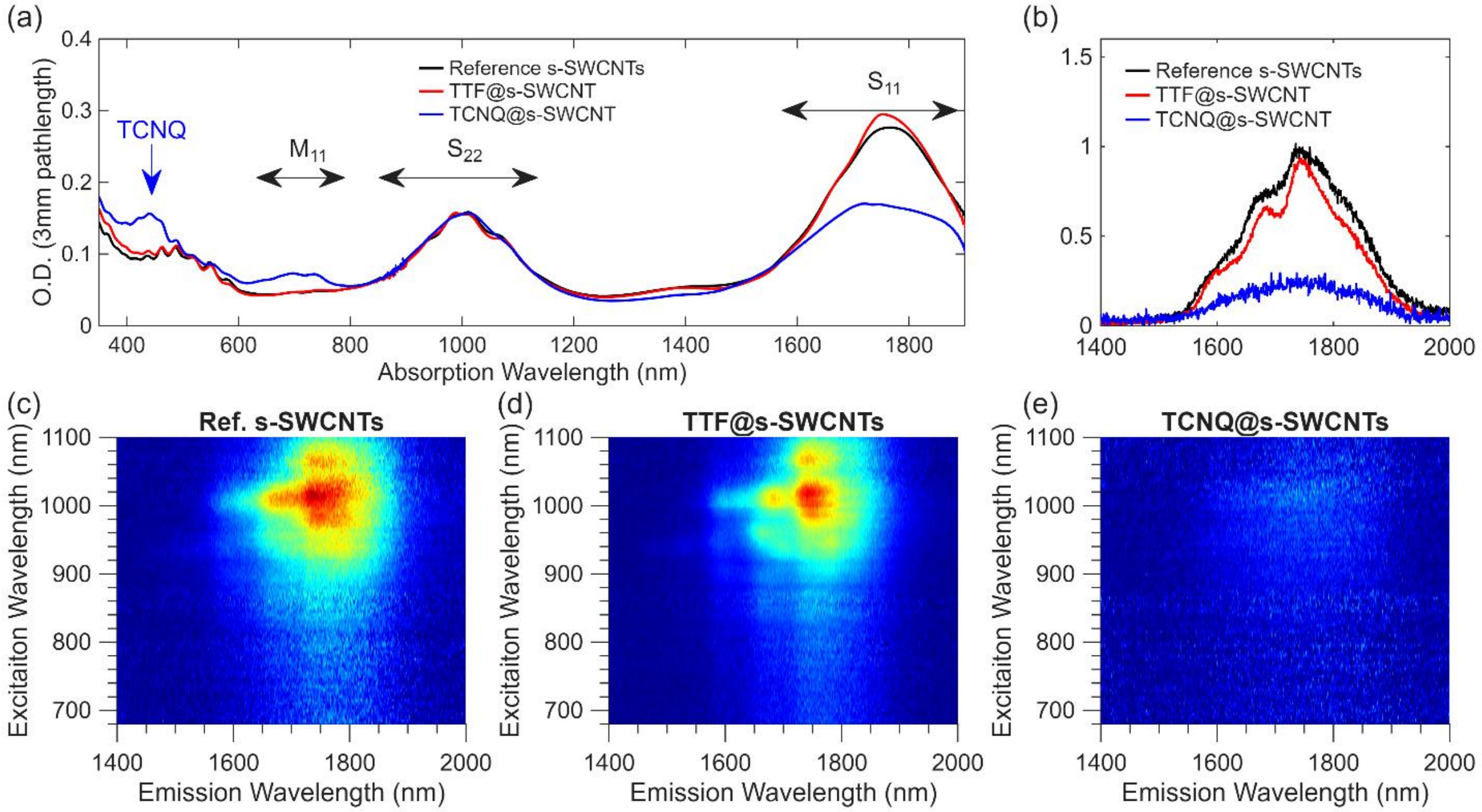

**Figure 11:** Metal-Semiconductor sorting of doped SWCNT samples. (a) absorption spectra of the semiconducting-enriched fractions after normalization on the $S_{22}$ resonance, showing the strong reduction of the $M_{11}$ peak. (b) Emission spectra, normalized over peak absorption, integrated over the PLE maps between 980-1030nm showing the differences in PL quantum efficiency and spectral properties and (c-e) PLE spectra of the three different samples.

### 3.8 XPS measurements to assess the degree of doping

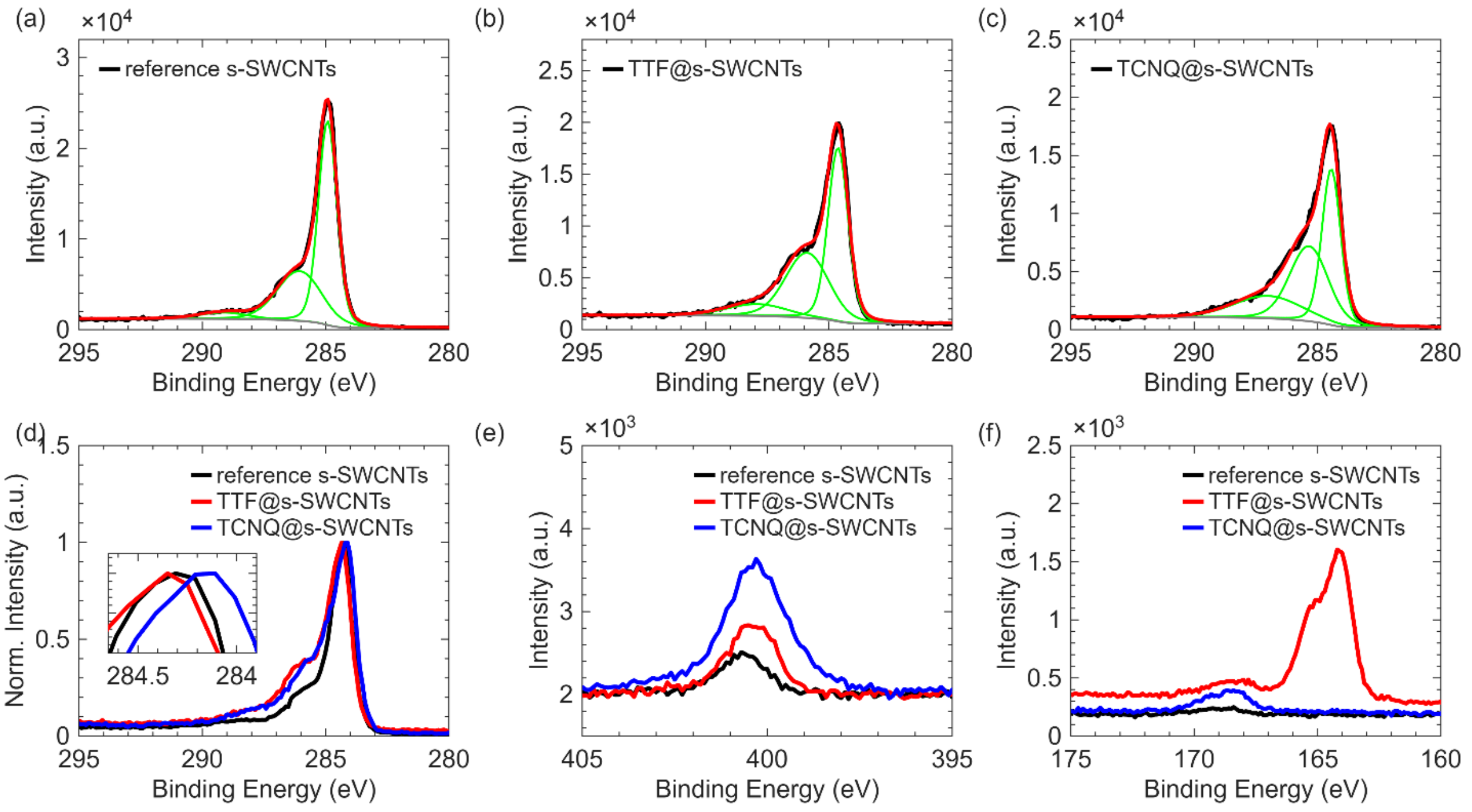

**Figure 12:** XPS data of s-SWCNT thin films (a-c) C 1s spectra (in black) for reference s-SWCNTs (a), TTF@s-SWCNTs (b) and TCNQ@s-SWCNT with superimposed fits (red) being a sum of three asymmetric Lorentzian functions (in green) and a Shirley background (in grey). The lowest energy peak corresponds to the sp$^2$ carbon. (d) comparison of the three XPS C1s spectra after normalization and calibration with the Si substrate peak, showing clear shifts of the doped SWCNTs with respect to the reference SWCNTs. (e) N 1s and (f) S 2p spectra of the three samples.

**Figure 12a-c** shows the fitted C 1s spectra of the three thin film samples, while similar results were obtained for the powdered samples right after FES filling (SI **Figure S10**). For the reference unfilled SWCNT sample, the main C 1s peak is located at 284.33 eV, which is consistent with the expected binding energy of sp$^2$ hybridized carbon,[54] particularly when considering the potential uncertainty associated with the energy calibration.

**Figure 12d** compares the C 1s spectra of the three samples after the correction with the Si peak from the substrate. Compared with the SWCNT reference sample, the TCNQ-filled CNTs exhibit a noticeable shift of the C 1s main peak towards lower binding energy by 0.18 eV, while the TTF-filled CNTs show a smaller shift towards higher binding energy by 0.05 eV. These trends are consistent with previous reports showing that p-doping may result in a downshift of the C 1s peak, whereas n-doping can lead to an upshift.[55–57] The relatively small upshift observed for the TTF- filled CNTs is in agreement with the weaker n-type doping also observed in optical spectroscopy, due to the competing effects of p-type doping from external adsorption of oxygen. **Figure 12e-f** show N 1s and S 2p spectra of the three samples. A clear sulfur signal is observed for the TTF($C_6H_4S_4$)-filled CNTs, while a pronounced nitrogen signal is detected for the TCNQ($C_{12}H_4N_4$)-filled CNTs. These signals confirm the presence of the respective filling molecules in the CNT films. It should be noted that nitrogen is also detected in the reference and TTF samples due to the $Si_3N_4$ substrate. However, the substantially stronger nitrogen signal observed for the TCNQ sample confirms an additional contribution from the encapsulated TCNQ molecules. In addition, XPS measurements were also performed on the powder samples of the three types. (see **Figure S10**). In these measurements a nitrogen signal was only observed in the TNCQ-filled CNTs confirming that the nitrogen signal in the other samples arose due to the $Si_3N_4$ substrates.

## CONCLUSIONS

In summary, we present a systematic comparative study of multiple encapsulation strategies for introducing electron-donor (TTF) and electron-acceptor (TCNQ) molecules into arc-discharge SWCNTs, covering the complete workflow from raw CNT powders to individually solubilized and metal/semiconductor-sorted dispersions. By combining Raman spectroscopy, optical absorption, and TGA, we demonstrate that vacuum-phase sublimation consistently provides the highest filling efficiency and the strongest electronic modification, while solution-reflux and melt filling yield lower encapsulation or introduce practical limitations for subsequent sample processing. We further show that conventional rinsing is insufficient to fully remove externally adsorbed dopants, particularly for the poorly soluble TCNQ, and that oxidative cleaning, though effective, undesirably degrades the nanotubes. To address this, we introduce filling and extraction by sublimation as a robust methodology enabling efficient filling while selectively removing surface-bound molecules under dynamic vacuum, without structural damage and without solvent-induced artifacts. EPR confirms the doping towards the SWCNT lattice by leaving behind oxidized TTF or reduced TCNQ. Importantly, we confirm that endohedrally doped SWCNTs remain fully compatible with post-processing steps such as density-gradient ultracentrifugation, allowing both empty–filled and metal–semiconductor sorting to be performed after doping. Moreover, XPS measurements confirm the direction of the doping. Overall, our results provide a quantitative framework for assessing encapsulation efficiency, identify the most reliable routes for stable n- and p-type molecular doping, and establish FES as a scalable and versatile strategy for preparing clean, processable, and electronically well-defined doped SWCNTs for future nanoelectronic and optoelectronic applications, such as transistors and logic devices.

**Acknowledgements**

The authors thank Horizon Europe under the EIC-Pathfinder project (3D-BRICKS, GA 101099125). C.B. also thanks the “Bijzonder Onderzoeks Fonds BOF” from the University of Antwerp for a one year PhD grant (FN 542300009). C.M.B-P, S.C. and K.Y. acknowledge funding for a JSPS/FWO joint mobility grant (JPJSBP120252302, JSPS Japan and VS02425N, FWO, Belgium) for funding a research stay of C.M.B.-P. to the group of K.Y. K.Y. acknowledges support from JST ASPIRE project, Grant No. JPMJAP2310 and JST CREST through Grant No. JPMJCR2544, Japan. W.W. acknowledges support from the Research Foundation – Flanders (FWO) through project no. G099424N. S.C. thanks the FWO for providing infrastructure funding for the EPR instrumentation (I004920N) and the University of Antwerp CASCH Centre of Excellence. S. F. acknowledges the French Research Agency through the ANR JCJC grant Dinosaures. T.S. and M.S. acknowledge the support of the RADIANT project (BIST IGNITE SEED 8th EDITION). T.S and M.S. also thank Guillaume Sauthier for his technical support with the XPS measurements.

**CRedIT**

**Cristian M. Borja-Peña:** Data Curation, Formal Analysis, Investigation, Methodology, Validation, Visualization, Writing – original draft

**Martin Magg:** Data Curation, Investigation, Methodology, Formal Analysis, Writing – review and editing

**Hussein Slim**: Data Curation, Investigation, Writing – review and editing

**Yamaldi M. Bakary**: Data Curation, Investigation, Writing – review and editing

**Timm Swoboda:** Data curation, Investigation, Methodology, Formal Analysis, Writing – review and editing

**Cedric Desroches**: Supervision, Writing – review and editing

**Denys Kurylo**: Data Curation, Writing – review and editing

**Kazuhiro Yanagi**: Supervision, Methodology, Writing – review and editing

**Marianna Sledzinska:** Supervision, Data curation, Investigation, Writing – review and editing

**Benjamin S. Flavel**: Supervision, Conceptualization, Methodology, Resources, Funding Acquisition, Writing- review and editing

**Salomé Forel:** Supervision, Conceptualization, Data Curation, Methodology, Resources, Writing- review and editing

**Wim Wenseleers:** Supervision, Conceptualization, Methodology, Resources, Funding Acquisition, Writing-review and editing

**Sofie Cambré**: Data Curation, Formal Analysis, Investigation, Methodology, Visualization, Supervision, Resources, Funding Acquisition, Conceptualization, Writing – original draft

**ORCID**

**Cristian M. Borja-Peña:** 0009-0005-6830-2923

**Martin Magg:** 0009-0005-9825-0511

**Hussein Slim**:

**Yamaldi M. Bakary**: 0009-0000-0346-7798

**Timm Swoboda:** 0009-0009-8072-4204

**Cedric Desroches**: 0000-0003-4433-5147

**Denys Kurylo**: 0009-0005-5506-9595

**Kazuhiro Yanagi**: 0000-0002-7609-1493

**Marianna Sledzinska:** 0000-0001-8592-1121

**Benjamin S. Flavel**: 0000-0002-8213-8673

**Salomé Forel:** 0000-0002-3706-3676

**Wim Wenseleers:** 0000-0002-3509-0945

**Sofie Cambré**: 0000-0001-7471-7678

**TOC Figure**

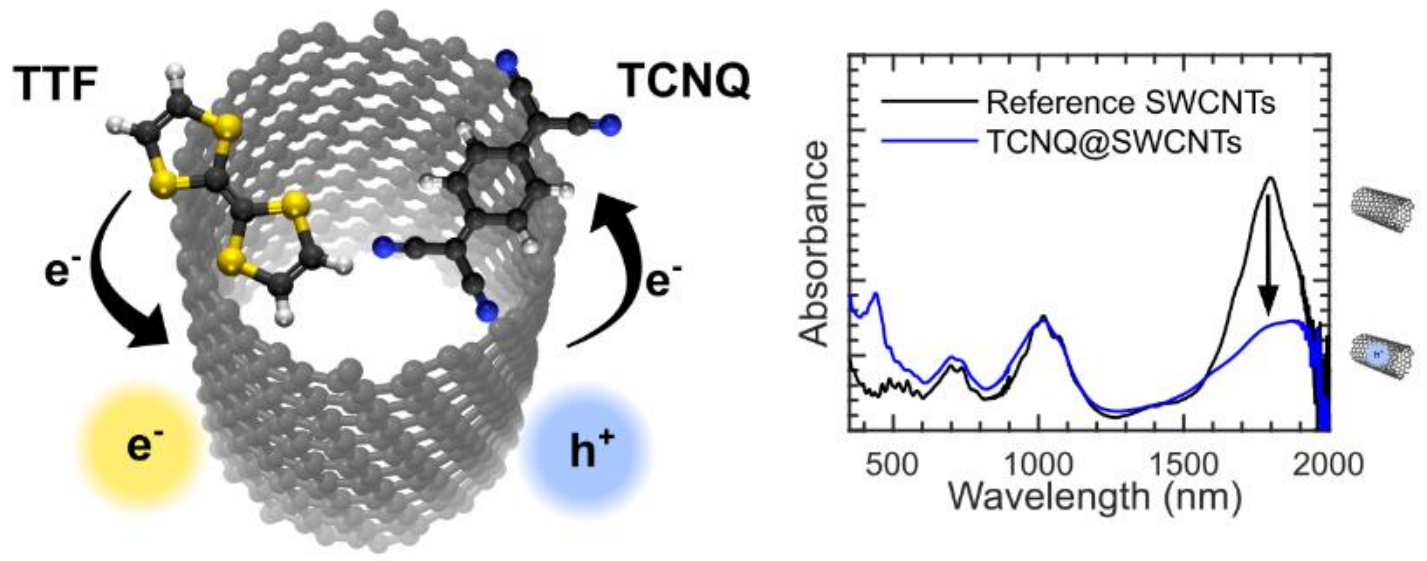

TTF
TCNQ
e-
e-
e-
h+
Reference SWCNTs
TCNQ@SWCNTs
Absorbance
500
1000
1500
2000
Wavelength (nm)

# Supplementary Materials for

# Comparative Evaluation of Encapsulation Methods for Endohedral Doping of Single-Wall Carbon Nanotubes

*Cristian M. Borja-Peña[1,2,3], Martin Magg[4], Hussein Slim[5], Yamaldi M. Bakary[5], Timm Swoboda[6], Cedric Desroches[5], Denys Kurylo[3], Kazuhiro Yanagi[7], Marianna Sledzinska[6], Benjamin S. Flavel[4], Salomé Forel[5,*], Wim Wenseleers[3], Sofie Cambré[1,2,*]*

[1] University of Antwerp, Physics Department, Theory and Spectroscopy of Molecules and Materials (TSM[2]).

[2] CASCH Center of Excellence, University of Antwerp, Antwerp, Belgium.

[3] University of Antwerp, Physics Department, Nanostructured and Organic Optical and Electronic Materials (NANOrOPT).

[4] Institute of Nanotechnology, Karlsruhe Institute of Technology, Kaiserstraße 12, 76131 Karlsruhe, Germany.

[5] Claude Bernard University Lyon 1, Laboratoire des Multimatériaux et Interfaces (LMI).

[6] Catalan Institute of Nanoscience and Nanotechnology (ICN2), CSIC and BIST, Barcelona, 08193, Spain.

[7] Tokyo Metropolitan University, Department of Physics, 1-1 Minami-Osawa, Hachioji, Tokyo, Japan 1920397.

**Section S1: Extinction coefficients and Solubility of TTF and TCNQ**

A suitable solvent for reflux-based filling must have (i) a low boiling point, (ii) a high enough solubility to allow an excess of molecules relative to the SWCNTs, and at the same time (iii) a low enough solubility to enable straightforward preparation of a saturated solution. To this end, methanol (MeOH) was identified as an effective solvent for TTF-based filling, while dichloromethane (DCM) was used for TCNQ.

First, the extinction of TTF and TCNQ was determined by preparing solutions of known concentration in MeOH and DCM, respectively, and measuring their absorption spectrum. From this the maximum extinction coefficient was extracted for both molecules (Figure S1a). Next, a saturated solution was prepared, and its absorption spectrum was recorded and the absorbance at the peak was divided by the corresponding extinction coefficient, allowing the solubility (saturated concentration) to be extracted (Figure S1b). From these measurements, a solubility of 0.045 mol/L or 9.2 g/L (Molar Mass 204.36g/mol) was obtained for TTF, consistent with the requirements outlined above. Similarly, for TCNQ with DCM as the most suitable solvent for rinsing of excess molecules, yields a solubility of 0.012 mol/L or 2.45 g/L (Molar Mass 204.19 g/mol).

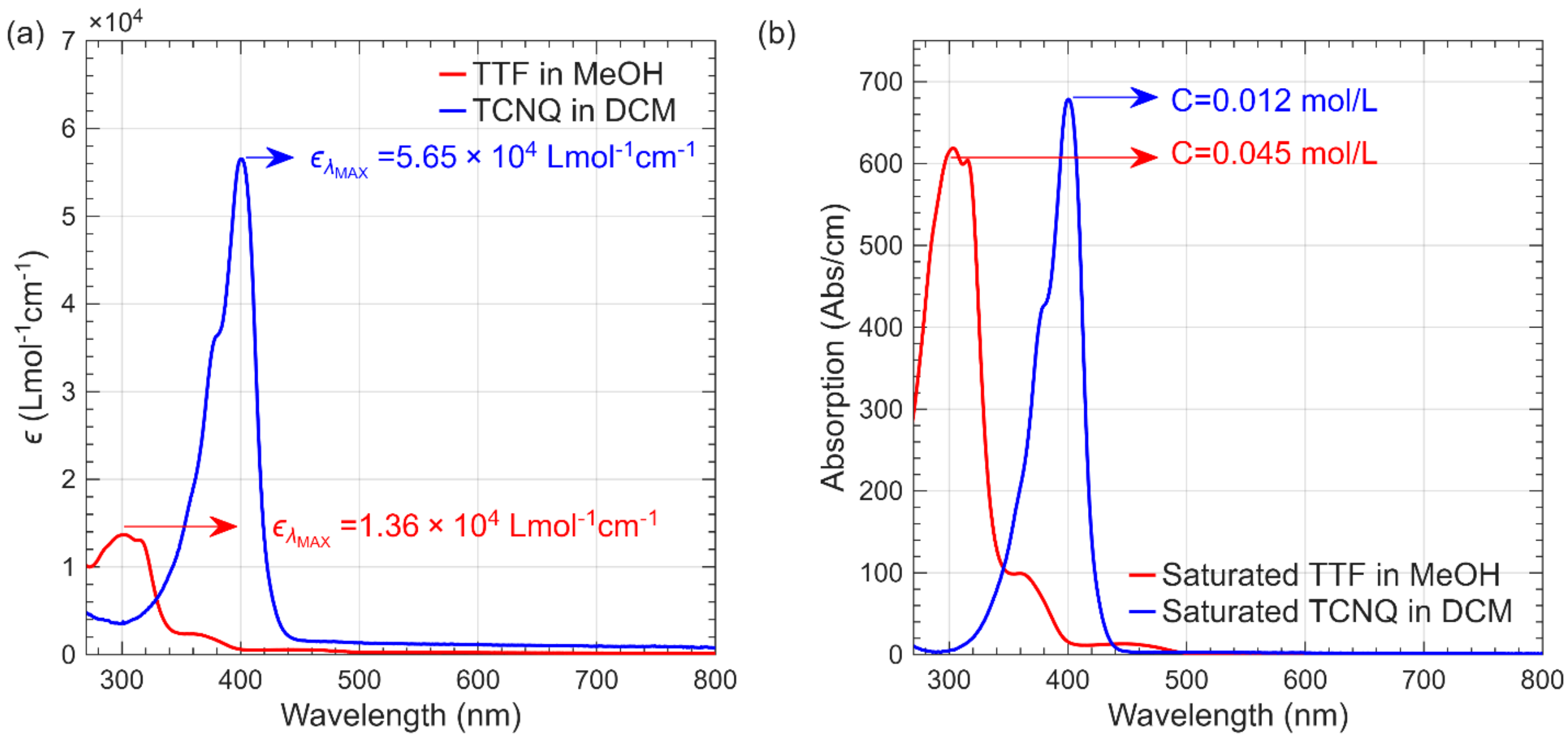

**Figure S1: (a)** Extinction coefficient of TTF in MeOH (red) and TCNQ in DCM (blue). The values in the graph represent the extinction coefficients at the maximum absorption wavelength. (b) Absorption spectrum of a saturated solution of TTF in MeOH (red) and TCNQ in DCM (blue) divided by the path length ($l$ = 1 cm). The concentration can then be read at this maximum absorption wavelength, being 300 nm for TTF and 400 nm for TCNQ.

**Section S2: Filling and extraction by sublimation (FES)**

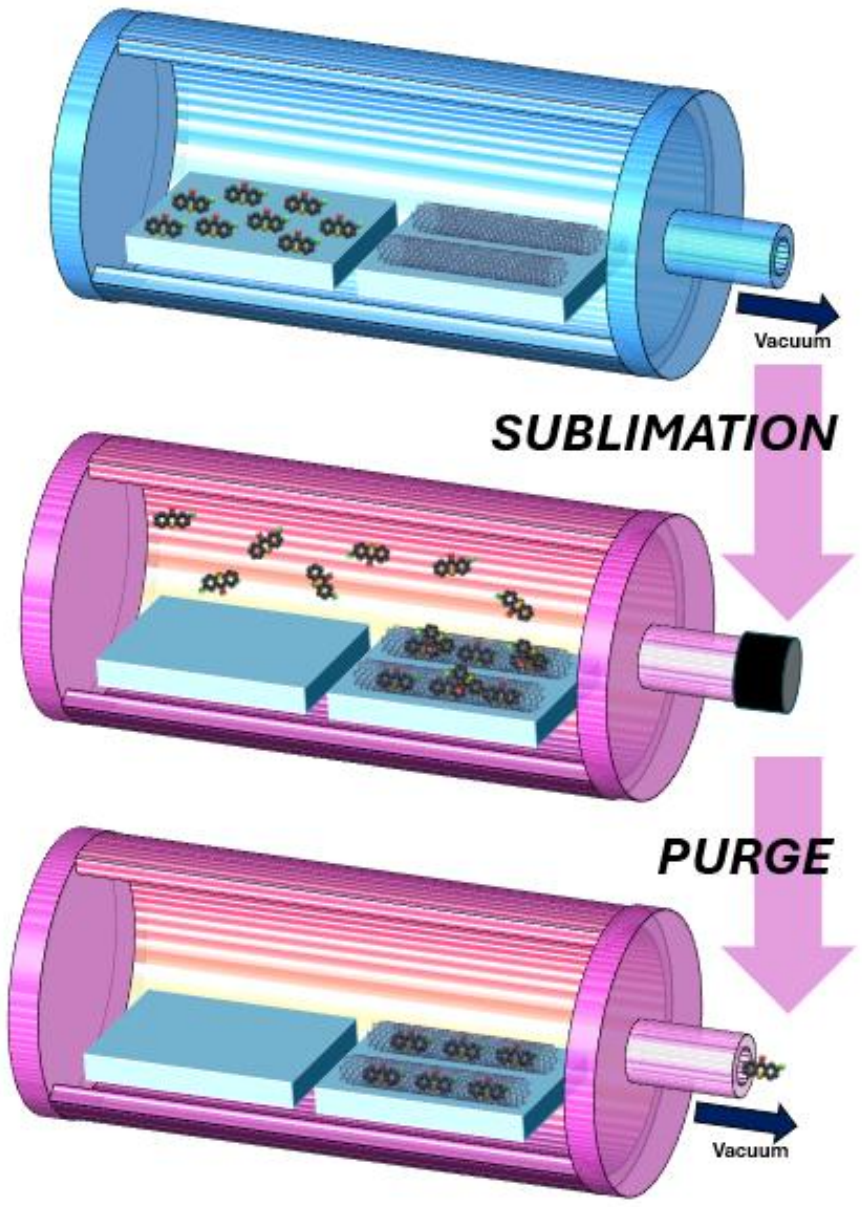

**Figure S2:** schematic setup for filling and extraction by sublimation (FES), including a filling step under static vacuum at increased temperature, and a removal step under dynamic vacuum at increased temperature.

**Table S1:** Conditions for FES of SWCNTs with TTF and TCNQ. The ones in bold have been discussed in more detail in the main text.

| Sample | Filling Temp | time | Removal Temp | Removal Time |
|---|---|---|---|---|
| **S1: TTF** | **250˚C** | **1h** | **250˚C** | **3h** |
| S2: TTF | 200˚C | 1h | 200˚C | 3h |
| S3: TTF | 200˚C | 3h | 200˚C | 3h |
| S4: TCNQ | 250˚C | 3h | 250˚C | 3h |
| **S5: TCNQ** | **250˚C** | **1h** | **250˚C** | **3h** |

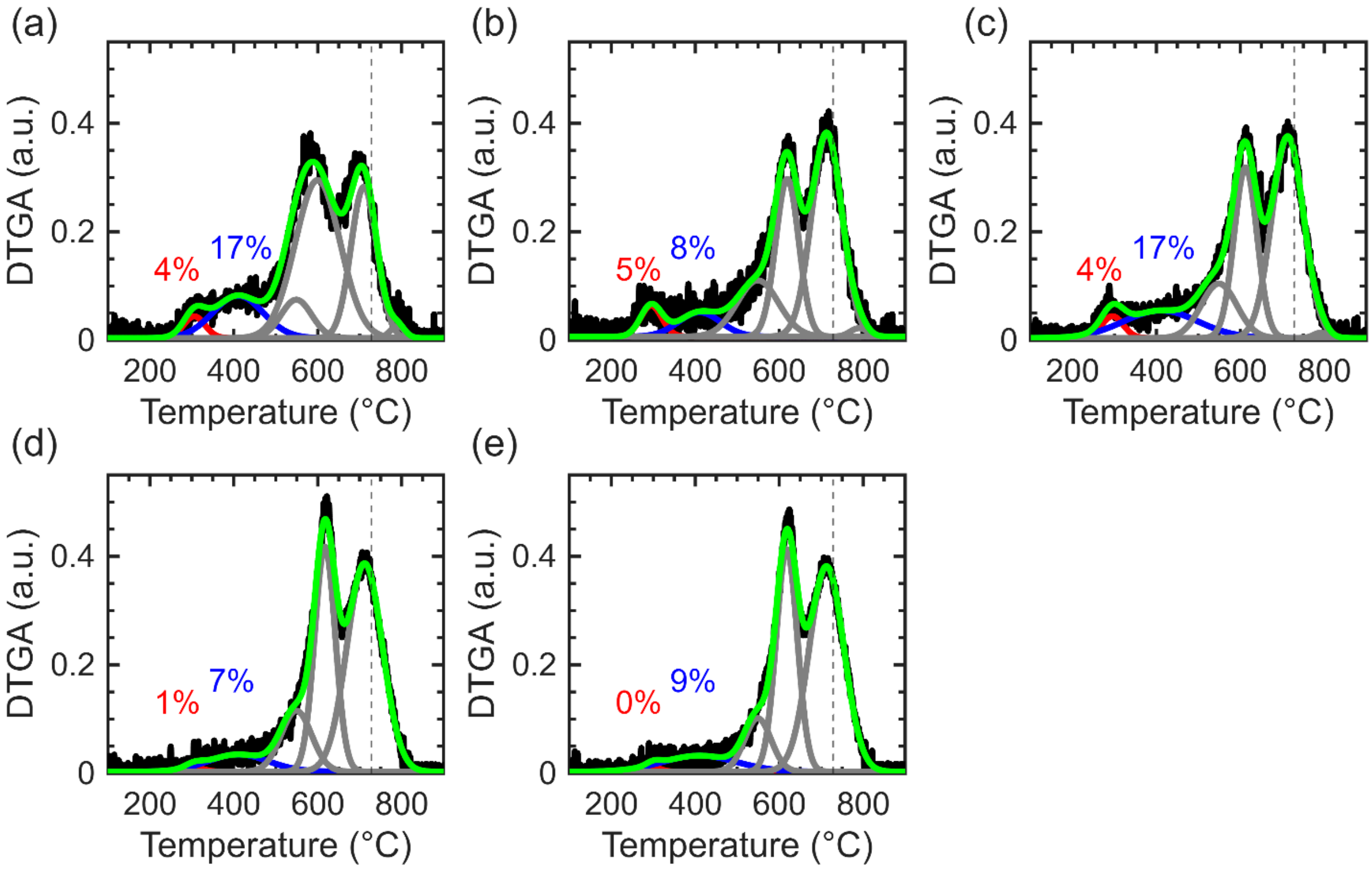

**Figure S3:** TGA analysis of the 5 samples prepared with the FES parameters as declared in Table S1. (a) S1, (b) S2, (c) S3, (d) S4, (e) S5. Samples S1 and S5 were used in the main text.

## Section S3: Rinsing of excess molecules

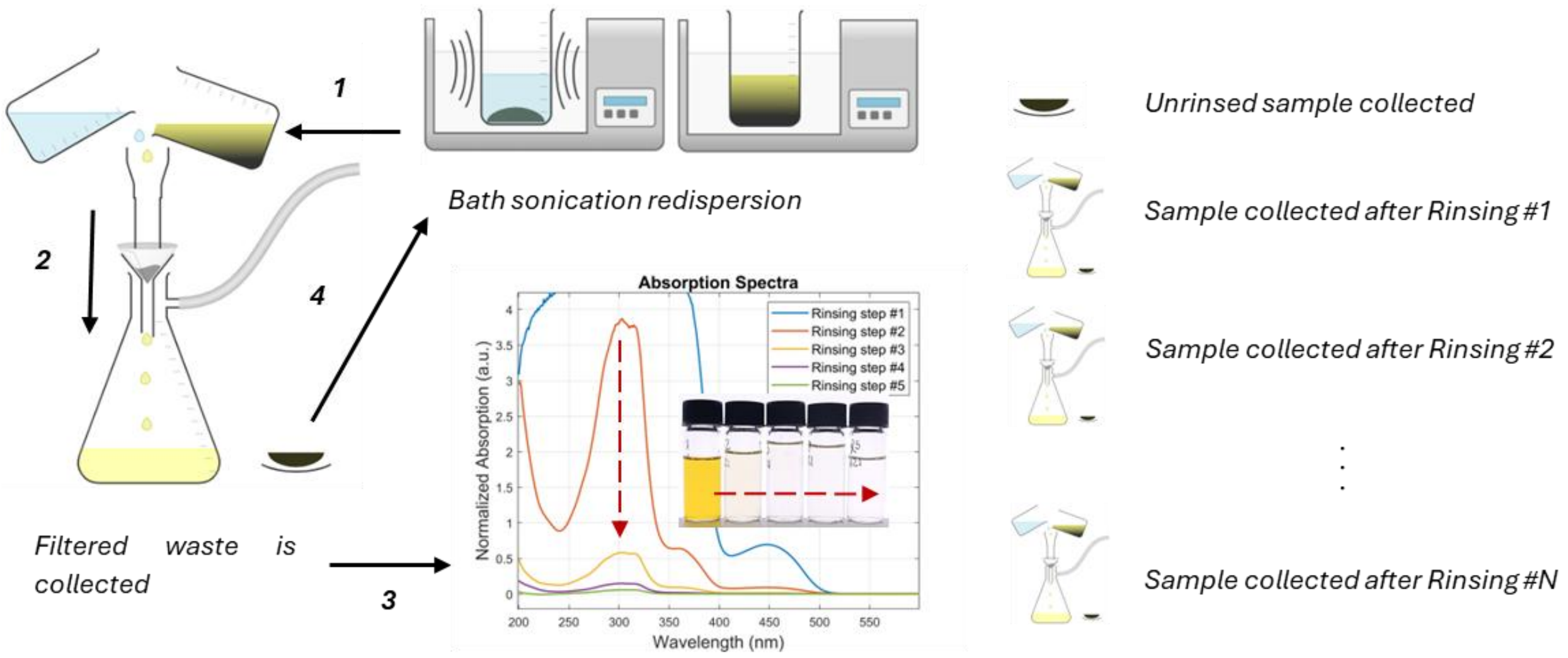

**Figure S4:** Schematic representation of the extensive rinsing methodology. (1) First, the SWCNTs are briefly sonicated in a solvent and (2) subsequently poured on top of a filter membrane. Then, pure solvent is used to extensively wash the filter membrane and sample. (3) The filtered waste is collected, and an absorption spectrum is taken for each rinsing step. (4) The SWCNTs on the filter membrane are redispersed in a solvent after which step (1) repeats itself. Note that in between each rinsing cycle, a small fraction of the CNT powder sample is collected for characterization. In total 5 rinsing steps were done.

**Section S3: DGU for empty-filled and metallic/semiconducting sorting**

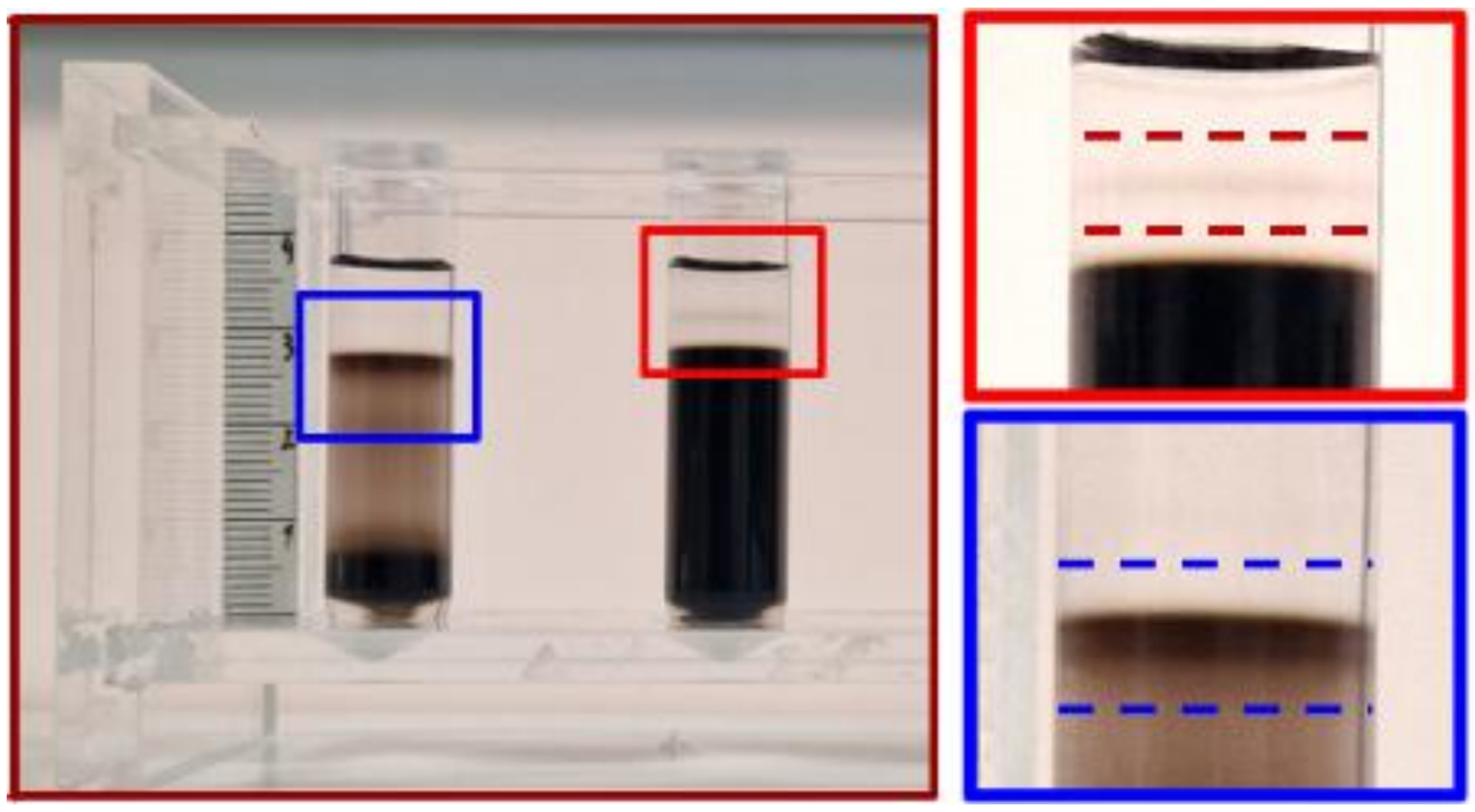

**Figure S5:** Sorting of empty (red lines) and filled (blue lines) SWCNTs through DGU.

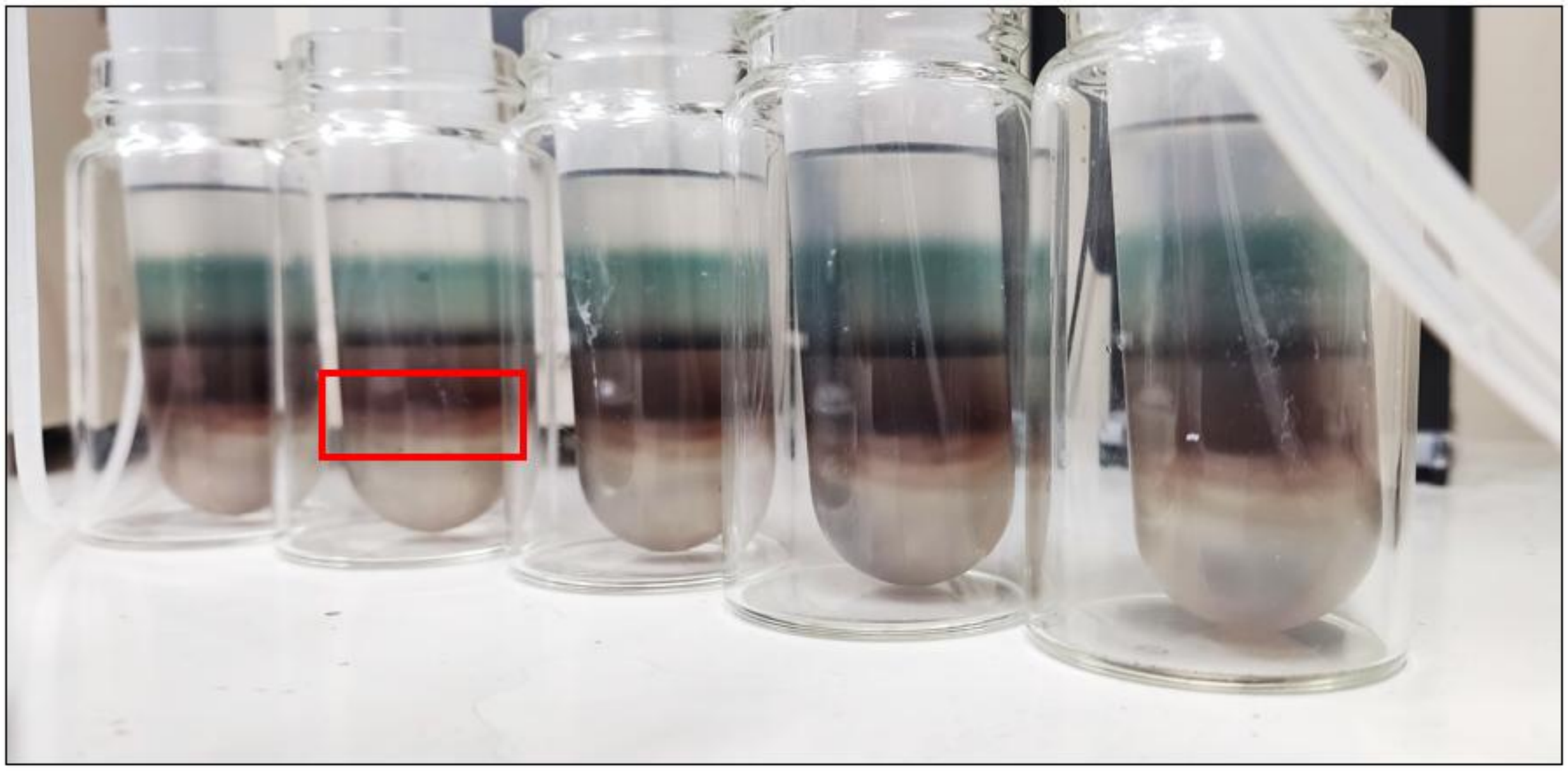

**Figure S6:** DGU for metallic-semiconducting sorting of SWCNTs. The red rectangle indicates the semiconducting fraction that was collected for further characterization.

**Section S4: Solubility of different TTF-filled samples**

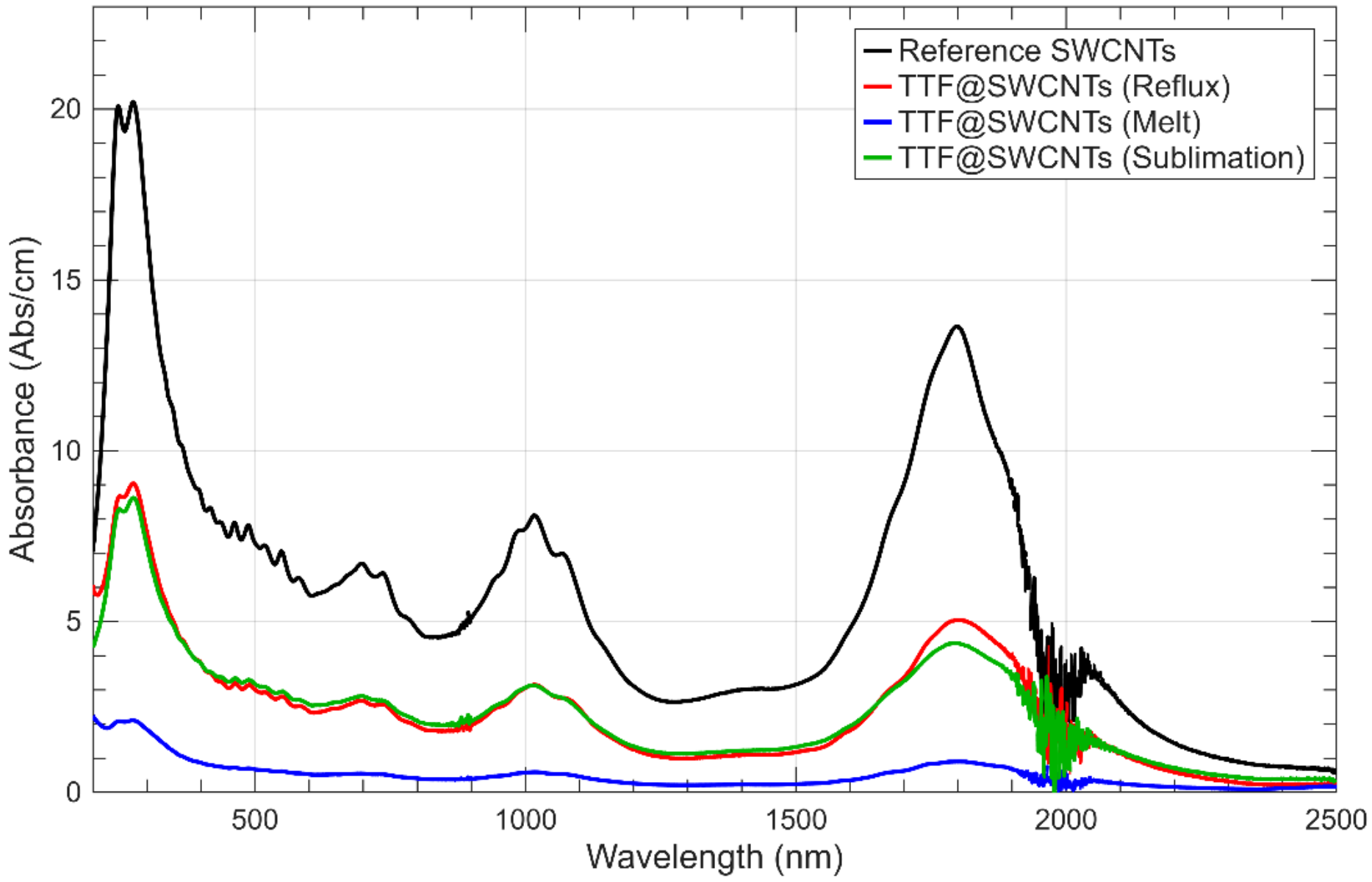

**Figure S7:** Absorbance (measured in a 3mm path-length cell after proper dilution and then rescaled according to this dilution factor) of the different samples, starting from the same amount of nanotube powder and using the exact same conditions for solubilization and centrifugation. From this it can be concluded that solubilization from the melt-phase filling is much more difficult as the sample is more compacted after filling, while reflux and sublimation give a very similar optical density.

**Section S5: XPS measurements and film formation**

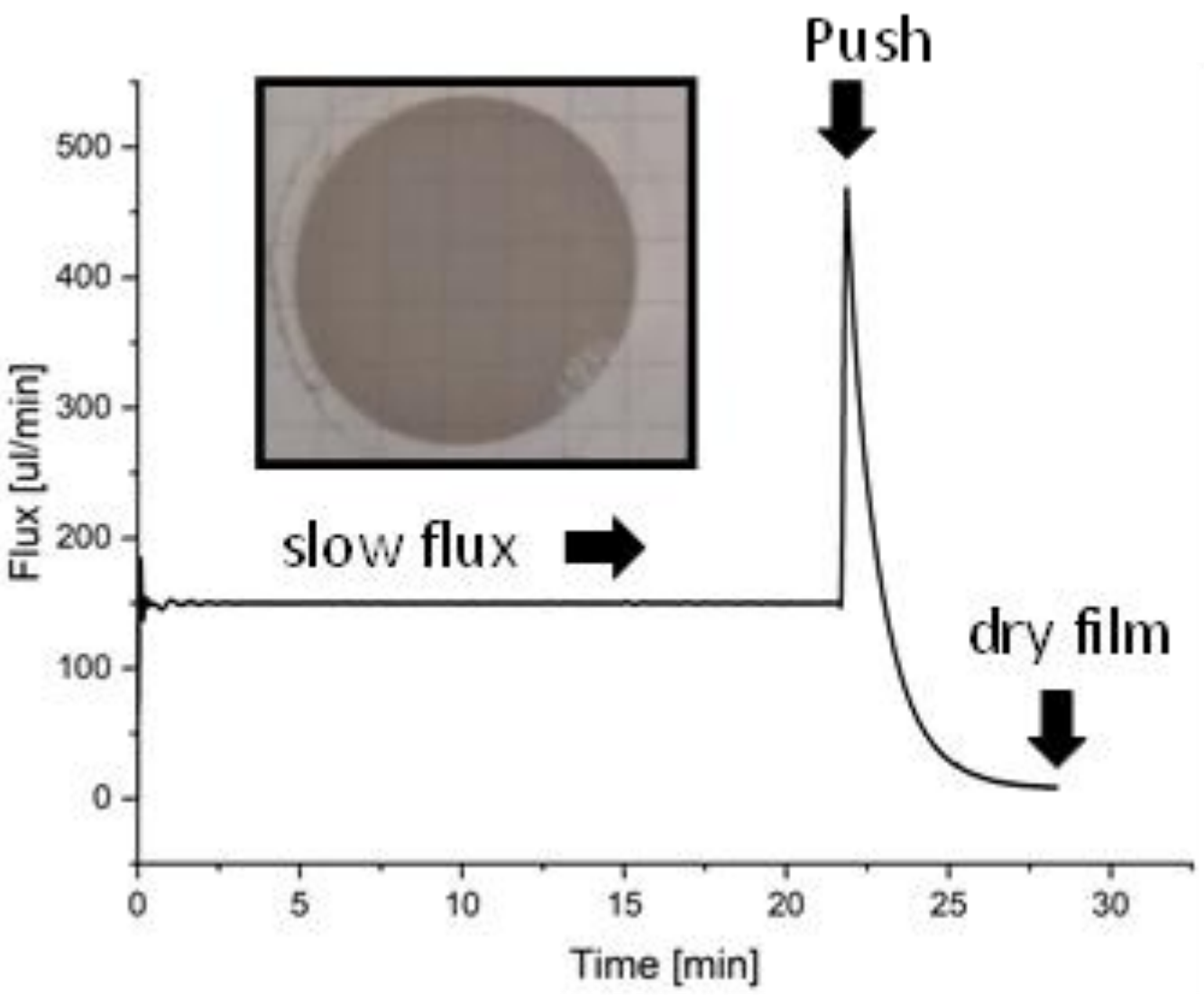

**Figure S8:** Typical profile of the flow as a function of time for a dead-end filtration of a s-SWCNT film. Initial filtration is performed at constant flow rate of 150 µl min$^{-1}$, which ensures steady deposition of nanotubes on the membrane to produce a dense film of randomly oriented SWCNTs. After 20 min (3250 µL) the flow is increased to 500 µL min$^{-1}$ to push material on to the membrane. The inset shows a photograph of a prepared film on a polycarbonate membrane.

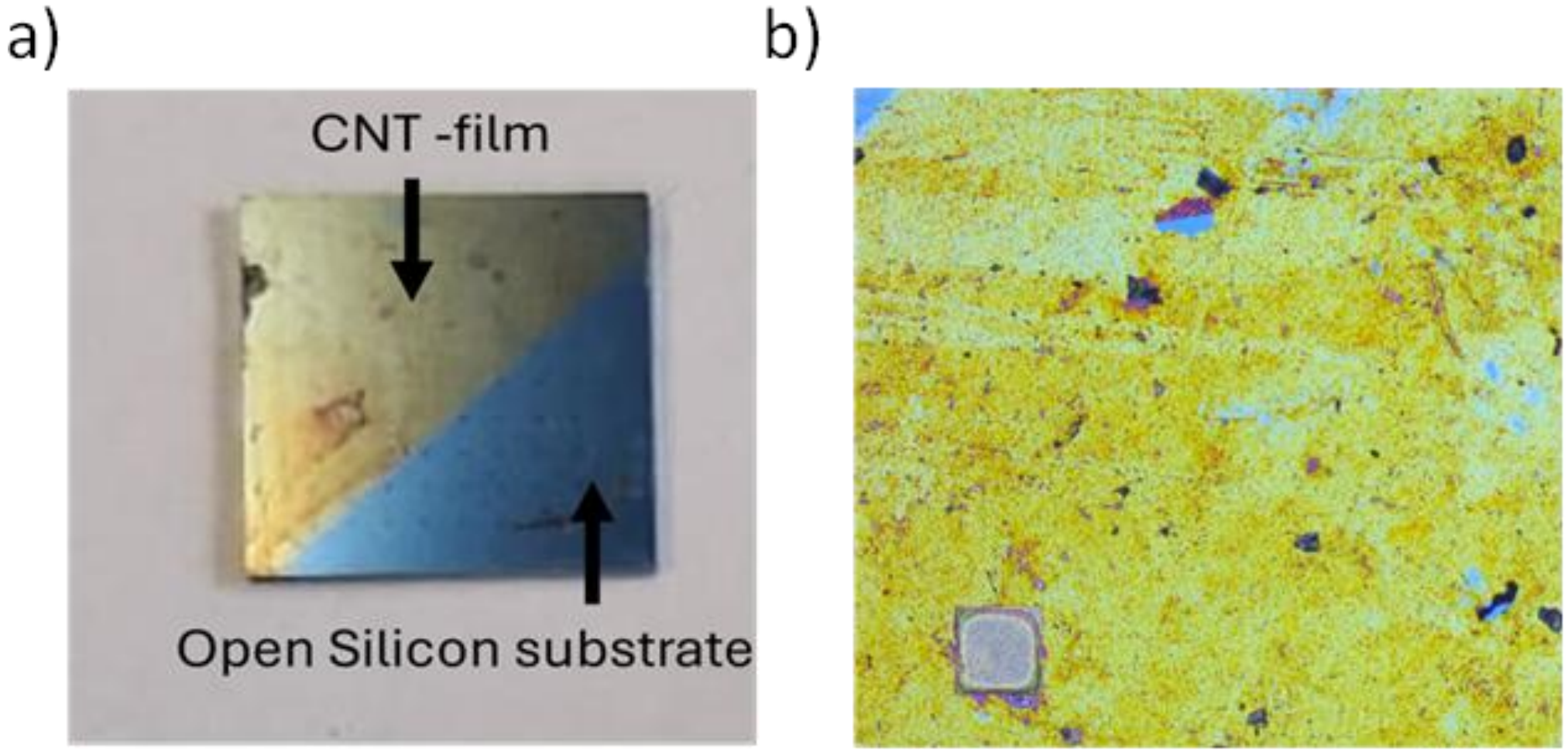

**Figure S9:** s-SWCNT films transferred to silicon nitride substrates. a) Photograph of a silicon nitride substrate covered and uncovered with a CNT-film b) bright-field microscopy image of a transferred SWCNT film. Bottom left corner: square opening with an edge length of 75 µm used as a scale bar.

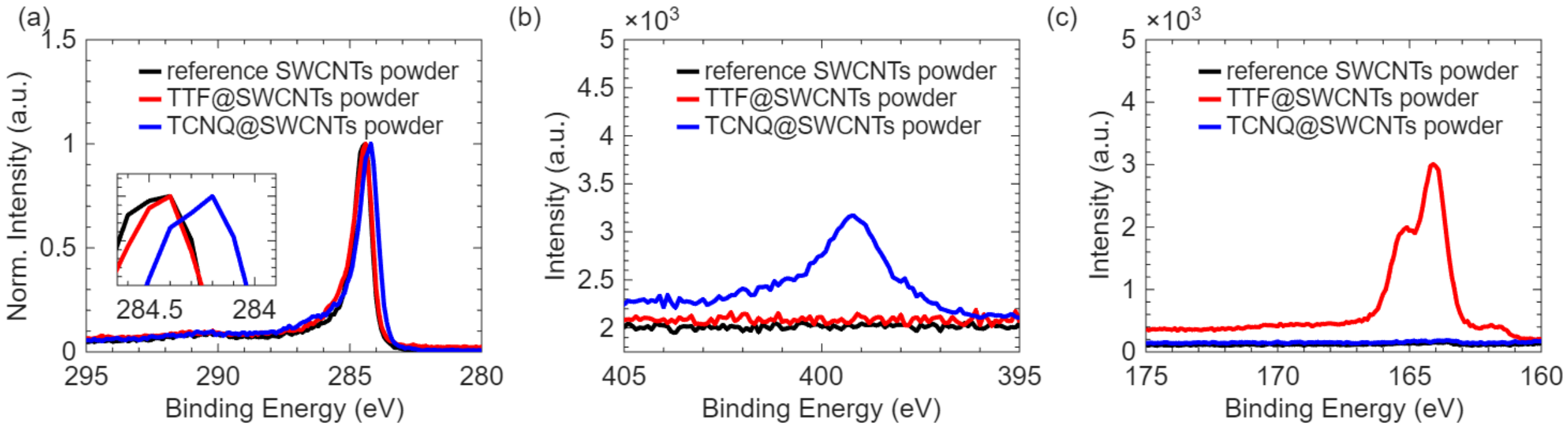

**Figure S10**: XPS measurements of the FES powders, directly after filling without any sorting. (a) C 1s spectra comparison of TTF, TCNQ and reference FES powders showing similar shifts as for the XPS s-SWCNT films. In this case, however, no calibration with Si could be performed as the powders were directly measured without the silicon nitride substrate (b) N 1s spectra showing only a contribution in the TCNQ sample and (c) S 2p spectra showing only a contribution for the TTF sample.